\documentclass[letterpaper]{JHEP3}

\usepackage{amssymb}
\usepackage{amsmath}
\usepackage{graphicx}
\usepackage{bm} 
\usepackage{amssymb}
\usepackage{amstext}
\usepackage{tensor}
\usepackage{epsf,epsfig}
\usepackage{amsfonts} 
\usepackage{graphicx}
 \usepackage{bm} 
 \usepackage{cite}

\usepackage{amsmath}
\usepackage{mathrsfs}
 
\usepackage{amsmath,amssymb}

\newcommand{\insertplot}[5]{\begin{figure}
 \hfill\hbox to 0.05in{\vbox to #5in{\vfill
 \inputplot{#1}{#4}{#5}}\hfill}
 \hfill\vspace{-.1in}
 \caption{#2}\label{#3}
 \end{figure}}
 \newcommand{\inputplot}[3]{
 \special{ps: plotfile #1}
}
\newcounter{fig}   

\newcommand{\ze}{\kern 0.05em}

\newcommand{\eps}{\epsilon}

\newcommand{\beq}{\begin{equation}}
\newcommand{\eeq}{\end{equation}}
\newcommand{\beqs}{\begin{eqnarray}}
\newcommand{\eeqs}{\end{eqnarray}}

\newcommand{\be}{\begin{equation}}
\newcommand{\ee}{\end{equation}}
\newcommand{\bea}{\begin{eqnarray}}
\newcommand{\eea}{\end{eqnarray}}

\newcommand{\identity}{{\upright\rlap{1}\kern 2.0pt 1}}

\def\theequation{\arabic{equation}}

\numberwithin{equation}{section}

\abstract{ 
Spherical black hole (BH) solutions in Einstein-Maxwell-scalar (EMS) 
models wherein the scalar field is non-minimally 
coupled to the Maxwell invariant by some coupling 
function are discussed. We suggest a classification
 for these models into two classes, 
based on the properties of the coupling function,
 which, in particular, allow, or not, the Reissner-Nordstr\"om (RN) BH 
solution of electrovacuum to solve a given model. 
Then, a comparative analysis of two illustrative 
families of solutions, one belonging to each class 
is performed: $dilatonic$ versus $scalarised$ BHs. 
 By including magnetic charge, that is considering dyons, 
we show that scalarised BHs can have a smooth extremal limit, 
unlike purely electric or magnetic solutions. 
In particular, we study this extremal limit using the entropy 
function formalism, which provides insight on why both charges are necessary for extremal solutions to exist.

}

\keywords{ black holes, numerical solutions, attractors}\preprint{ }

\title{ 
Einstein-Maxwell-scalar black holes:  \\ classes of solutions, dyons and extremality
}

\author{{\large } 
{\large D. Astefanesei}$^{\dagger}$,
{\large C. Herdeiro}$^{\ddagger}$,
{\large A. Pombo}$^{\star}$
and {\large E. Radu}$^{\star}$
\\ \\
$^{\dagger}${\small 
Pontificia Universidad Cat\'olica de
	Valpara\'\i so, Instituto de F\'{\i}sica
	},
 \\ 
{\small Av. Brasil 2950, Valpara\'{\i}so, Chile}
\\
$^{\ddagger}${\small CENTRA,  Departamento  de  F\'\i sica,  Instituto  Superior  T\'ecnico  -  IST,
}
\\
{\small
Universidade  de  Lisboa  -  UL,  Avenida  Rovisco  Pais  1,  1049  Lisboa,  Portugal}
$^{\star}${\small Departamento de F\'isica da Universidade de Aveiro and CIDMA,
}
\\
{\small
 Campus de Santiago, 3810-183 Aveiro, Portugal}

 }

\begin{document}
 
\newpage
  
\section{Introduction}
Einstein-Maxwell (EM) theory, \textit{a.k.a.} electrovacuum, is the quintessential source-free, gravitational relativistic field theory. Its static physical black holes (BHs) belong to the 3-parameter Reissner-Nordstr\"om (RN) family, described by mass $M$, electric $Q$ and magnetic $P$ charges. These BHs are perturbatively stable~\cite{Moncrief:1974gw,Moncrief:1974ng}, and, for given $M$, can only sustain charges $(P,Q)$ if $\sqrt{Q^2+P^2}\leqslant M$. When the equality holds, the extremal limit is attained. Extremal RN BHs are special. They are non-singular spacetimes, on and outside a degenerate and $C^\infty$ smooth event horizon, that: (i) have a vanishing Hawking temperature and are BPS states that possess Killing spinors when embedded in supergravity~\cite{Gibbons:1982fy}; (ii) have a near horizon geometry which is, itself, a solution of the EM theory~\cite{Gibbons:1993sv} -- the Robinson-Bertotti ($AdS_2\times S^2$) vacuum~\cite{Robinson:1959ev,Bertotti:1959pf}; and (iii) allow a no-force condition and a multi-BH generalisation, described by the Majumdar-Papapetrou metrics~\cite{Majumdar:1947eu,Papapetrou:1948jw,Hartle:1972ya}. 

A simple and natural generalisation of the EM theory is to consider an additional dynamical real scalar field, with a standard kinetic term. A variety of such EM-scalar (EMS) models are possible, depending on the way the scalar field couples to the Maxwell field.\footnote{Herein we shall always consider that  the scalar field is minimally coupled to gravity.}  Remarkably, if the scalar field is minimally coupled to the Maxwell, no new charged BH solutions are possible,  beyond RN, even if the scalar field is allowed to have a non-negative self-interactions potential~\cite{Mayo:1996mv}. In these conditions, charged BHs cannot have scalar \textit{hair}. Quite different possibilities, however, arise if a \textit{non-minimal} coupling between the scalar and Maxwell field is allowed. This is the case we shall be interested in this paper.

The first such non-minimally coupled EMS model emerged in the pioneering unification theory of Kaluza 
\cite{Kaluza:1921tu}
and Klein
\cite{Klein:1926tv},
soon after Einstein constructed General Relativity (GR)~\cite{Einstein:1915ca}. 
These EMS models turned out to be ubiquitous in the four dimensional description of higher dimensional
GR-inspired theories~\cite{Appelquist:1987nr}, as well as in supergravity, see $e.g.$~\cite{Cremmer:1978ds}.
 In the former, as well as in the latter with a higher dimensional origin, the scalar field describes how the extra dimension(s) dilate along the four dimensional spacetime, being dubbed $dilaton$. The dilaton has a specific non-minimal coupling with the Maxwell term in the EMS action. This coupling prevents EM theory to be a consistent truncation of this class of EM-dilaton models. 
In particular, the RN solution of EM theory does not solve these EM-dilaton models. 
 Instead, new charged BHs with a non-trivial scalar field profile exist~\cite{Gibbons:1987ps,Garfinkle:1990qj}, which are known in closed analytic form and that present RN-unlike features.
For instance, the BH charge to mass ratio can exceed unity (see, $e.g.$~\cite{Delgado:2016zxv}). As another example, there are no  extremal BHs with a regular horizon in the purely electric (or purely magnetic) 
case.
 These limiting solutions 
become naked singularities\footnote{Asymptotically flat, purely electric BHs exist also 
in EMS models with a non-trivial scalar potential,
explicit solutions being reported in 
\cite{Anabalon:2013qua, Astefanesei:2019mds}.
Since
in this case there are 
two different terms that source the scalar field 
(a self-interaction potential and the term coming from the non-trivial coupling with the electric field), 
a balance is possible, which allows for a well defined extremal limit 
\cite{Anabalon:2013sra}.}, a sharp contrast with the physically interesting extremal RN BH. Nonetheless, dilatonic BHs provided an example of asymptotically flat charged BHs with scalar hair~\cite{Gibbons:1982ih}, albeit of secondary type~\cite{Herdeiro:2015waa}.

 Once embedded in string theory, the dilaton $\phi$ controls the string coupling, which
 is related to the vacuum expectation of the asymptotic value of the dilaton, $g_s=e^{\langle\phi_\infty \rangle}$. Therefore, a consistent analysis of hairy BHs in string theory should consider a dynamical dilaton whose asymptotic value can vary \cite{Gibbons:1996af} (see, also, \cite{Astefanesei:2018vga} for a resolution of the appearance of the scalar charges in the first law of thermodynamics). This need, however, is mitigated by \textit{the attractor mechanism} \cite{Ferrara:1995ih, Ferrara:1996dd}: the near horizon data (particularly, the entropy) of extremal BHs is independent of the asymptotic values of the moduli. The mechanism is based on a simple physical intuition; when the temperature vanishes, there is a symmetry enhanced near horizon geometry: $AdS_2\times S^2$. The infinite long throat of $AdS_2$ yields the decoupling between the physics at the boundary from the physics at the extremal horizon \cite{Kallosh:2006bt}. A similar decoupling plays a central role in the $AdS/CFT$ duality (see, $e.g.$  \cite{Larsen:2018iou, Lin:2019qwu}). 

EMS models with more generic non-minimal couplings (than the dilatonic one) between the scalar field and the Maxwell term are also of interest.
For example, such models were considered in cosmological inflationary scenarios~\cite{Martin:2007ue,Maleknejad:2012fw}. In the context of BHs, it was recently realised that a family of couplings can trigger a \textit{spontaneous scalarisation} of the RN BH~\cite{Herdeiro:2018wub,Fernandes:2019rez}. In this class of EMS models, unlike the aforementioned dilatonic models, the RN BH is a solution. For sufficiently large charge to mass ratio, however, the RN BH becomes unstable against scalar perturbations and \textit{dynamically} grows a scalar field profile;  it becomes energetically favourable to scalarise. The hair growth stalls due to non-linear effects leading to a scalarised BH (to be distinguished from dilatonic BH). The fundamental scalarised charged BHs, which are the ones formed dynamically are, moreover, perturbatively  stable~\cite{Myung:2018vug,Myung:2018jvi,Myung:2019oua} and therefore represent the endpoint of the non-linear evolution of the unstable RN BHs. Consequently, these scalarised BHs are an example of \textit{dynamically} grown scalar hair.

The scalarised BHs studied up to now contain only electric charge. They possess no extremal limit.
 Rather, a critical solution is attained for the maximal charge a BH can support, which (numerical evidence suggests) is singular. This parallels the status of dilatonic BHs. For the latter, however, the introduction of an \textit{additional} magnetic charge leads to dyonic BHs with an extremal (non-singular) limit, which have been constructed for specific couplings~\cite{Dobiasch:1981vh,Gibbons:1985ac,Kallosh:1992ii}. Given the importance of extremal solutions, it is interesting to inquire which are the properties of the family of dyonic scalarised BHs and, in particular, of their extremal limit.

 In fact, the considerations above suggest a comparison between dilatonic and scalarised BHs 
can be instructive. The purpose of this paper is to perform such a comparison, for the  canonical dilatonic coupling and the reference model of scalarised solutions introduced in~\cite{Herdeiro:2018wub}. Our results, within this comparative study, include: $(i)$ the introduction of a general framework to study EMS for any scalar non-minimal coupling; $(ii)$ the first study of dyonic scalarised BHs; $(iii)$ establishing that extremal scalarised BHs indeed exist (only) when both electric and magnetic charges are present; and $(iv)$ the study of the corresponding near horizon geometries via the attractor and entropy function formalism of~\cite{Sen:2005wa, Astefanesei:2006dd, Sen:2007qy}.

This paper is organised as follows. In Section~\ref{section2} we present the EMS models and propose a classification of the BH solutions, based on the behaviour of the coupling function. We also derive the zero mode of the RN BHs for the models that allow BH scalarisation.
Section~\ref{sec3} contains a discussion of the
 non-extremal BHs  for both two classes of solutions (dilatonic and scalarised).
In Section~\ref{sec4}, we study the extremal BHs toghether with the corresponding near horizon geometries, using the attractors formalism.
We conclude in Section~\ref{sec5} with a discussion and some further remarks.
The Appendix contains a brief review of the known exact solutions, all of which occur for a dilatonic coupling.

\section{The EMS model}
\label{section2}

\subsection{The action and equations of motion}

The EMS family of models is defined by the following action
(we set $c=G=4\pi \epsilon_0=1$) 
\begin{eqnarray}
\label{action}
\mathcal{S}= \frac{1}{16 \pi}\int d^4 x \sqrt{-g} 
\left(R-2\partial_\mu \phi\partial^\mu \phi 
- f(\phi) F_{\mu\nu}F^{\mu\nu} \right),
\end{eqnarray}
where $R$ is the Ricci scalar, $F_{\mu \nu}=\partial_\mu A_\nu-\partial_\nu A_\mu$ is the Maxwell field 
and $\phi$ is the scalar field. 
The coupling function $f(\phi)$ governs the non-minimal coupling of $\phi$ to the electromagnetic field. From the outset we are excluding an axion-type coupling of the scalar to the electromagnetic field, 
as well as any sort of self-interaction of the scalar field. 
The model may, of course, be generalised in these directions.  

The field equations obtained by varying the above action principle
with respect
to the field variables $g_{\mu \nu}$, $\phi$ and $A_{\mu}$ are
\begin{eqnarray}
\label{eqEinstein}
R_{\mu\nu} - \frac{1}{2}Rg_{\mu\nu} &=& 
2\left[\partial_{\mu} \phi\partial_{\nu}\phi 
-\frac{1}{2}g_{\mu\nu}\partial_{\rho}\phi\partial^{\rho}\phi 
+f(\phi) \left(F_{\mu\rho}F_{\nu}^{~\rho} 
- \frac{1}{4}g_{\mu\nu} F_{\rho\sigma}
F^{\rho\sigma} 
\right)
  \right] \ ,
\\
\label{eqScalar}
\frac{1}{\sqrt{-g}}\partial_{\mu}(\sqrt{-g}\partial^{\mu}\phi) 
&=&  \frac{1}{4}\frac{df(\phi)}{d\phi}
F_{\rho\sigma}F^{\rho\sigma} \   ,
\\ 
\label{eqM}
\partial_{\mu}(\sqrt{-g}f(\phi)F^{\mu\nu})&=& 0 \ .
\end{eqnarray}

An ansatz suitable to address both the (generic)
asymptotically flat solutions and the
Robinson-Bertotti (near horizon) geometries reads
\begin{eqnarray}
\label{metric-generic}
 ds^2=-a(r)^2 dt^2 + b(r)^2 (d\theta^2+\sin^2\theta d\varphi^2) +c(r)^2dr^2 \ .
\end{eqnarray}
The gauge 4-potential ansatz compatible with the symmetries of (\ref{metric-generic}) contains an electric potential $V(r)$
and a magnetic term, 
\begin{eqnarray}
\label{A}
A=V(r) dt+P \cos \theta d\varphi~,
\end{eqnarray}
where $P$=constant  is the magnetic charge. The  
scalar field is a function of $r$ only, $\phi \equiv \phi(r)$.

The Maxwell equation~\eqref{eqM} yields a first integral
\begin{eqnarray}
\label{first-int}
V'=  \frac{ a c}{b^2 } \frac{Q}{f(\phi)} \ ,
\end{eqnarray}
where $Q$=constant is the electric charge (measured at infinity), and henceforth a prime denotes a derivative $w.r.t.$
the radial coordinate $r$.

The equations of motion~\eqref{eqEinstein}-\eqref{eqM} are invariant under the \textit{electro-magnetic duality} transformation
\begin{eqnarray}
\label{transf}
\{P\to Q,~~Q\to P \}~~{\rm and}~~f(\phi)\to 1/f(\phi) \ .
\end{eqnarray}
In what follows, we shall assume, without any loss of generality,
that both $Q$ and $P$ are positive and that 
\begin{eqnarray}
\label{qp}
Q\geqslant P \ ,
\end{eqnarray}
such that for scalarised BHs,
the (electric) solutions in \cite{Herdeiro:2018wub,Fernandes:2019rez}  are recovered as  $P\to 0$.

\subsection{The coupling function and a classification of EMS models}
The RN BH is a solution  of~\eqref{eqEinstein}-\eqref{eqM} with $f(\phi)=1$, $\phi$=constant and
\begin{eqnarray}
\label{RN}
V(r)=-\frac{Q}{r}\ , \qquad a(r)^2=\frac{1}{c(r)^2}=1-\frac{2M}{r}+\frac{Q^2+P^2}{r^2}\ , \qquad b(r)=r \ , 
\end{eqnarray}
where $M$ is the BH ADM mass.
For a more general $f(\phi)$ the RN BH may or may not solve~\eqref{eqEinstein}-\eqref{eqM}. This naturally leads to two classes of EMS models. (Note that, in this classification, we assume, without any loss of generality, that the scalar field vanishes asymptotically, $\phi(r)\stackrel{r\rightarrow \infty}\longrightarrow 0$.) 
\begin{description}
\item[Class I or dilatonic-type.] In this class of EMS models $\phi(r)= 0$ does $not$ solve the field equations.\footnote{There is an exceptional case: if $Q=P$, $\phi=0$ solves this class, so that the dyonic, equal charges RN BH is a solution.} 
 Thus RN is not a solution. 
Then, the scalar field equation (\ref{eqScalar})
 implies that 
\begin{eqnarray}
\label{condx}
f_{,\phi}(0)\equiv \frac{d f (\phi)}{d \phi}\Big |_{\phi=0} \neq 0\ .
\end{eqnarray}
A representative example of coupling for this class is the standard dilatonic coupling
\begin{eqnarray}
\label{dilaton}
f (\phi)=e^{2\alpha \phi} \ ,
\end{eqnarray}
in which case we refer to $\phi$ is a dilaton field. The 
arbitrary nonzero constant   $\alpha$ is taken to be positive without any
loss of generality. Indeed, the solutions remain invariant
under the simultaneous sign change $(\alpha,\phi) \to -(\alpha,\phi)$. Thus, flipping the sign of $\alpha$ simply corresponds to flipping the sign of $\phi$. The coupling~\eqref{dilaton} appears naturally in Kaluza-Klein models and supergravity/low-energy string theory models. Three reference values for the coupling constant $\alpha$ in~\eqref{dilaton} are:
\begin{equation}
\alpha=0 \ \ {\rm (EM \ theory)} \qquad \alpha=1 \ \ {\rm (low \ energy \ strings)} \qquad \alpha=\sqrt{3} \  \ {\rm (KK \ theory)} \ .
\end{equation}
Some exact, closed form BH solutions of~\eqref{eqEinstein}-\eqref{eqM} with~\eqref{dilaton} are known and presented in Appendix A. Other exact solution examples in this class (with a non-dilatonic coupling) are given in~\cite{Fan:2015oca}.
\item[Class II or scalarised-type.]  In this class of EMS models $\phi(r)= 0$ solves the field equations. Thus RN is a solution. This demands that 
\begin{equation}
f_{,\phi}(0)\equiv \frac{d f (\phi)}{d \phi}\Big |_{\phi=0}= 0 \ .
\label{typeii}
\end{equation}
This condition is naturally implemented, for instance, if one requires the model to be $\mathbb{Z}_2$-invariant under $\phi\rightarrow -\phi$. The RN solution, however, is (in general) not  unique. These EMS models may contain a second set of BH solutions, 
with a nontrivial scalar field profile -- {\it the scalarised BHs}. Below some conditions for this to occur are discussed. Such second set of BH solutions may, or may not, continuously connect with RN BHs. This leads to two subclasses.
\begin{description}
\item[Subclass IIA or scalarised-connected-type.] In this subclass of EMS models, the scalarised BHs bifurcate from RN BHs, and reduce to the latter for $\phi=0$. This bifurcation moreover, may be associated to a tachyonic instability, against scalar perturbations, of the RN BH. Considering a small-$\phi$ expansion of the coupling function
\begin{eqnarray}
f(\phi)=f(0)+\frac{1}{2}\frac{d^2 f(\phi)}{d \phi^2}\Big |_{\phi=0}+\dots \ ,
\end{eqnarray}
equation (\ref{eqScalar}) linearised for  small-$\phi$ reads:
\begin{eqnarray}
\label{eq-phi-small}
(\Box-\mu_{\rm eff}^2)\phi =0\ , \qquad {\rm where} \ \ 
\mu_{\rm eff}^2= \frac{F_{\mu\nu}F^{\mu\nu}}{4} \frac{d^2 f(\phi)}{d \phi^2}\Big |_{\phi=0}   \ .
\end{eqnarray}
The instability arises if $\mu_{\rm eff}^2<0$, which in particular requires
\begin{equation}
f_{,\phi\phi}(0)\equiv \frac{d^2 f(\phi)}{d \phi^2}\Big |_{\phi=0} \neq 0 \ ,
\end{equation}
and  with the opposite sign of $F_{\mu\nu}F^{\mu\nu}$. A reference example of a coupling function in this subclass, which we consider in this work is~\cite{Herdeiro:2018wub}
\begin{eqnarray}
\label{quadratic}
f (\phi)=e^{2\alpha \phi^2} \ ,
\end{eqnarray}
a case which is also relevant in cosmology
 \cite{Martin:2007ue,Maleknejad:2012fw}. Depending on the coupling, this subclass could also contain \textit{another} family of disconnected scalarised BHs, akin to the ones of class IIB below.
 \item[Subclass IIB or scalarised-disconnected-type.] 
 In this subclass of EMS models, the scalarised BHs do not bifurcate from RN BHs, and do not reduce to the latter for $\phi=0$. This is the case if there is no tachyonic instability of RN, for which a sufficient (but not necessary) condition is that 
\begin{equation}
f_{,\phi\phi}(0)\equiv \frac{d^2 f(\phi)}{d \phi^2}\Big |_{\phi=0} = 0 \ .
\end{equation}
We shall not address further this case in this paper (which, moreover, was not considered yet in the literature), 
but a representative coupling would be, say, 
$f(\phi)=1+\alpha \phi^4$.
\end{description}
\end{description}

Condition~\eqref{typeii} guarantees RN is a solution. But it does not guarantee the existence of scalarised BHs. In the case of purely electric (or magnetic) BHs, two  Bekenstein-type identities can be derived,  which put some constraints on $f(\phi)$ so that scalarised solutions exist. These can be derived as follows.  

To derive the first identity, the scalar field equation 
(\ref{eqScalar}) is multiplied  by $f_{,\phi}$ and integrated over a spacetime volume.
Integrating by parts and discarding the boundary terms, by virtue of the horizon properties and asymptotic flatness, one obtains
\begin{eqnarray}
\label{eq1}
 \int d^4 x \sqrt{-g} 
\left(
f_{,\phi \phi} \partial_\mu \phi \partial^\mu \phi+\frac{f_{,\phi}^2}{4}
F^2
\right)=0 \ .
\end{eqnarray}
The sign of neither term is fixed, in general, and specific considerations are required. For instance, a purely electric field has  $F^2<0$; this implies 
\begin{eqnarray}
\label{cond1}
f_{,\phi \phi} >0 \ ,
\end{eqnarray}
must hold for some range of the radial coordinate $r$, otherwise the two terms
of the integrand in (\ref{eq1}) will have always the same sign, making the identity only possible for $\phi=0$.

A second identity is found by multiplying (\ref{eqScalar})
by $\phi$,
which results, via a similar procedure, in 
\begin{eqnarray}
\label{eq2}
 \int d^4 x \sqrt{-g} 
\left(
  \partial_\mu \phi \partial^\mu \phi+\frac{\phi f_{,\phi}}{4} 
F^2
\right)=0 \ .
\end{eqnarray}
This imples that for 
a purely electric field  
the potential should satisfy the condition
\begin{eqnarray}
\label{condf}
\phi f_{,\phi}>0 \ ,
\end{eqnarray}
for some range of $r$.
Similar arguments hold for purely magnetic solutions, which implies $f_{,\phi \phi} <0$
and $\phi f_{,\phi}<0$, 
respectively.
No such results can be established in the generic dyonic case, since the sign of $F^2=F_{\mu \nu}F^{\mu \nu}$ is not determined,  \textit{a priori}.

\subsection{Spontaneous scalarisation of dyonic RN BHs: zero modes}
 Class IIA of EMS models is particularly interesting because it accommodates the dynamical phenomenon of spontaneous scalarisation~\cite{Herdeiro:2018wub,Fernandes:2019rez}
(see also
\cite{Gubser:2005ih},
\cite{Stefanov:2007eq},
\cite{Doneva:2010ke}
 for earlier discussions of charged BHs
scalarisation in different models). At the linear level this is manifest in the tachyonic instability~\eqref{eq-phi-small}. For a dyonic RN BHs (\ref{RN}),  $F_{\mu\nu}F^{\mu\nu}=-2(Q^2-P^2)/r^4$. Thus, under the assumption~\eqref{qp} a tachyonic instability requires $f_{,\phi\phi}(0)>0 $. Let us study this instability, generalising the analysis in~\cite{Herdeiro:2018wub,Fernandes:2019rez} for the dyonic RN case.

Assuming separation of variables,
\begin{eqnarray}
\label{p1}
 \phi=Y_{\ell m}(\theta,\varphi)U(r) \ ,
\end{eqnarray}
where $Y_{\ell m}$ are the real spherical harmonics and $\ell,m$ are
the associated quantum numbers, $i.e.$ $\ell=0,1,\dots$ and $-\ell\leqslant m \leqslant \ell$,
the  equation for the radial amplitude $U(r)$
reads
\begin{eqnarray}
\label{p2}
\left(\frac{r^2 U'}{c(r)^2}\right)'= 
\left[\ell(\ell+1)+\frac{ (P^2- Q^2)}{2r^2} f_{,\phi\phi}(0)
\right]U \ ,
\end{eqnarray}
where $c(r)$ is given by~\eqref{RN}.
Observe that the  term $ \mu^2_{\rm eff}=(P^2- Q^2)f_{,\phi\phi}(0)/r^2$  acts as the \textit{effective} mass 
  for the perturbations and the condition $\mu_{\rm eff}^2<0$ requires $f_{,\phi\phi}(0)>0 $, as discussed above.

For spherically symmetric perturbations $\ell=0$, and eq. (\ref{p2}) possesses an exact solution
which is regular on and outside the horizon and vanishes
at infinity\footnote{The limit  $P=0$ of this solution has been discussed in \cite{Fernandes:2019rez}.}
\begin{eqnarray}
\label{ex1}
U(r)=P_u 
\left(
1+\frac{2(Q^2-P^2)(r-r_H)}{r_H^2+P^2-Q^2}
\right)\ , \qquad {\rm where} \ \ u=\frac{1}{2}\left(\sqrt{1-2f_{,\phi\phi}(0)}-1\right) \ ,
\end{eqnarray}
where $r_H$ is the event horizon radial coordinate and $P_u$ is a Legendre function.
This solution is physical for $f_{,\phi\phi}(0)>1/2$, 
a condition which, for the coupling function (\ref{quadratic})  implies
\begin{eqnarray}
\alpha>1/8 \ .
\end{eqnarray}

For generic
parameters
 $(f_{,\phi\phi}(0),Q,P,r_H)$, 
the function
$U(r)$ 
approaches a constant \textit{non zero} value as $r\to \infty$,
\begin{equation}
\label{ex2}
U(r) \to U_\infty=
{}_2F_1 
\left[
\frac{1-\sqrt{1-2f_{,\phi\phi}(0)}}{2},\frac{1+\sqrt{1-2f_{,\phi\phi}(0)}}{2},1; \frac{Q^2-P^2}{Q^2-P^2-r_H^2}
\right]+\mathcal{O}\left(\frac{1}{r}\right).
\end{equation}
Thus finding the $\ell=0$ unstable mode
of a RN BH with given $P,Q,M$
reduces to a study of the zeros of the hypergeometric function
${}_2F_1 $, so that $U_\infty=0$.

\begin{figure}[ht!]
\begin{center}
\includegraphics[height=.34\textwidth, angle =0 ]{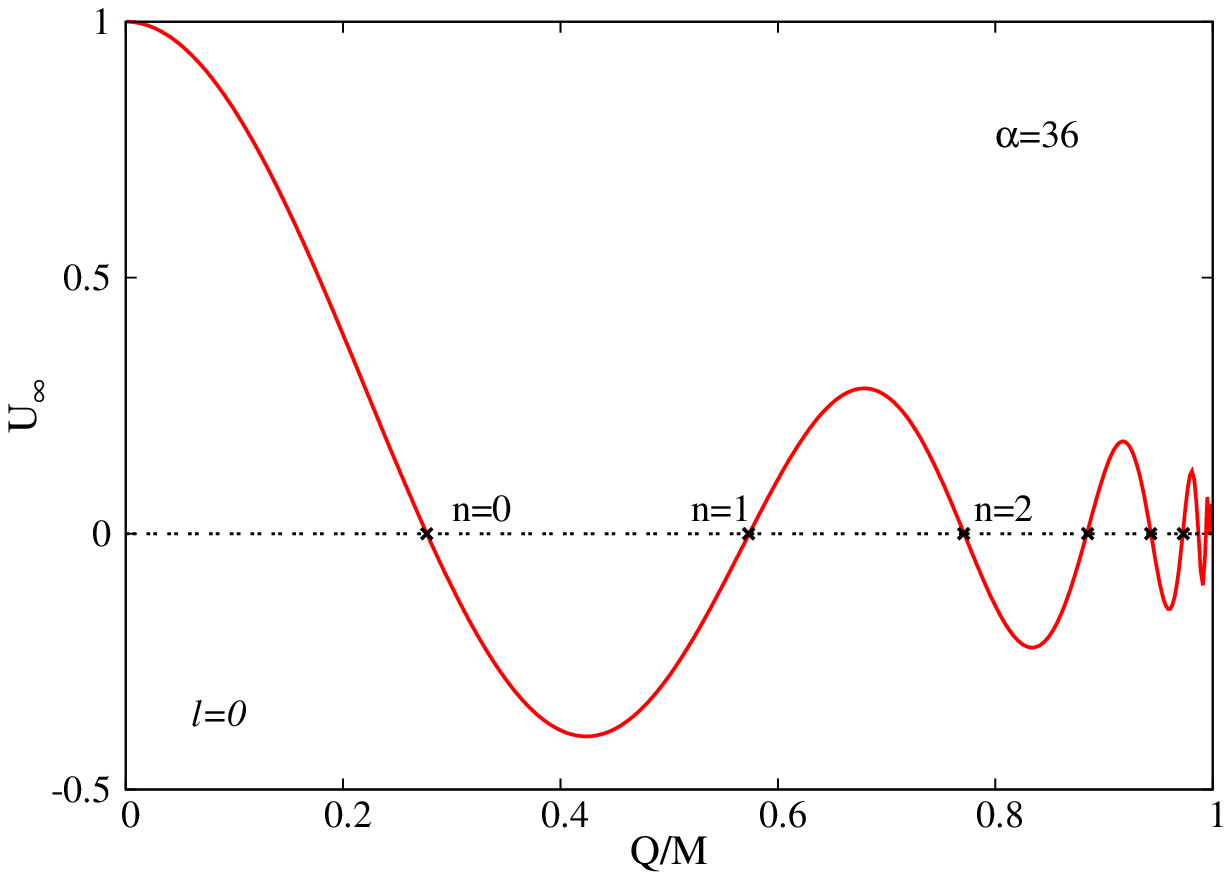}
\includegraphics[height=.34\textwidth, angle =0 ]{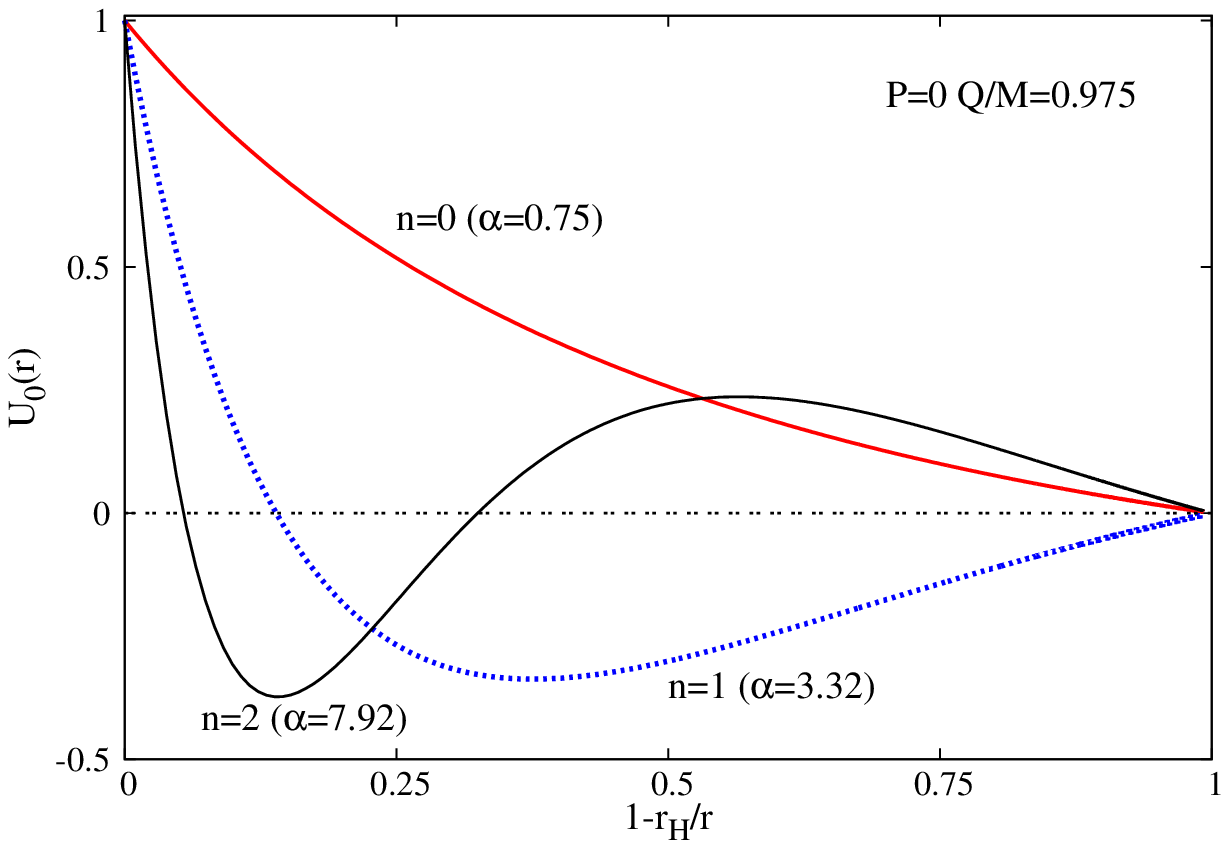}
\end{center}
\caption{
(Left panel)
 The asymptotic value  $U_\infty$ of the zero-mode amplitude $U$ for $\alpha=36$ as a function of the charge to mass ratio of a RN BH.
An infinite set of configurations with $U_\infty=0$ exist, labelled by $n$, the number of nodes of $U(r)$. 
(Right panel) The profiles of three zero mode amplitudes $U(r)$ with a different node number, for
a given RN background. 
}
\label{zero-mode}
\end{figure}

The value of $U_\infty$ for the coupling function (\ref{quadratic})  and an illustrative value of $\alpha$
is shown in Fig. \ref{zero-mode} (left panel).
Therein, the integer $n$ labels the number of nodes of the function $U(r)$ when the correct boundary condition at infinity is met:
given a RN background, the solutions with  $U_\infty=0$
are found for a discrete sequence  $\alpha_n$, each one corresponding to a different node number, $cf.$
Fig.~\ref{zero-mode} (right panel).
To simplify the picture,
the results  in Fig. \ref{zero-mode} correspond to $P=0$;  a similar pattern holds also in the dyonic case.
%

The solution  (\ref{ex1}) yields a dyonic RN BH surrounded by a vanishingly small scalar field. 
The set of
such RN BHs (varying $\alpha$) constitutes the {\it existence line}, the branching line between the RN and scalarised BHs. The latter are the non-linear continuation of the (infinitesimally small) scalar clouds, $i.e.$ the solutions of eq.~\eqref{p2}.
As remarked above, these clouds are labelled by three integer numbers $(\ell,m,n)$.
In what follows, however, we shall restrict our study to the simplest case of
nodeless, spherically symmetric configurations. More general configurations in the purely electric case have been discussed in~\cite{Herdeiro:2018wub,Fernandes:2019rez}. 

\section{Non-extremal black holes}
\label{sec3}
Let us now construct, numerically, the non-linear BH solutions, for both class I and IIA, starting with the non-extremal BHs.

\subsection{The ansatz and field equations}

In the numerical study of the solutions,
it is convenient to
work in Schwarzschild-like coordinates, with a metric gauge 
choice $b(r)=r$,  $a(r)^2=e^{-2\delta(r)}N(r)$ and $c(r)^2=1/N(r)$. Then,~\eqref{metric-generic} becomes:
\begin{equation}
\label{metric}
 ds^2=-e^{-2\delta(r)}N(r) dt^2+\frac{dr^2}{N(r)}+r^2 (d\theta^2+\sin^2\theta d\varphi^2)\ , \qquad {\rm where} \ \
N(r)\equiv 1-\frac{2m(r)}{r} \ .
\end{equation}
The function $m(r)$ corresponding to a local mass function, known as the Misner-Sharp mass~\cite{Misner:1964je}.

The  equations of motion  
(\ref{eqEinstein}) and (\ref{eqScalar}), 
together with the first integral (\ref{first-int})
implies that the functions $m$, $\sigma,\phi$
solve the ordinary differential equations
\begin{eqnarray}
\label{ec-m}
&&
m'=\frac{1}{2} r^2 N \phi'^2+\frac{1}{2r^2}\left(\frac{Q^2}{f(\phi)}+f(\phi)P^2 \right) \ ,
\\
\label{ec-d}
&&
\delta'+r \phi'^2=0\ , 
\\
&&
\label{ec-phi}
 (e^{-\delta} r^2 N\phi')'+\frac{ e^{-\delta}}{2r^2 f(\phi)} 
 \frac{df(\phi)}{d\phi} \left(\frac{Q^2}{f(\phi)}-f(\phi)P^2 \right)=0 \ ,
\end{eqnarray}
which can also be derived from  the following
effective action:
\begin{eqnarray}
\label{Seff}
S_{\rm eff}=\int dt dr ~
\left[e^{-\delta}m'-\frac{1}{2}e^{-\delta}r^2 N \phi'^2+\frac{f(\phi)}{2}\left(r^2 e^{\delta} V'^2-\frac{e^{-\delta}}{r^2}P^2\right)
\right] \ ,
\end{eqnarray}
while $V'=e^{-\delta}Q/(r^2 f(\phi))$.
The Einstein equations also yield a constraint  equation,
\begin{eqnarray}
\label{constr}
\frac{1}{2}N''-N \delta''
+N'\left(\frac{1}{r}-\frac{3}{2}\delta'\right)
+N\delta'\left(\delta'-\frac{1}{r}\right)
+N\phi'^2
-\frac{1}{r^4}
\left[
\frac{Q^2}{f(\phi)}+P^2 f(\phi)
\right]
=0 \ ,
\end{eqnarray}
 which can be shown to be a linear combination of equations
(\ref{ec-m})-(\ref{ec-phi}) together with the first derivatives of (\ref{ec-m})-(\ref{ec-d}).
 It is also of interest to observe that equations (\ref{ec-m})-(\ref{ec-phi}) 
possess the  first integral
\begin{eqnarray}
\label{Eqs-FistInt}
 e^{-2\delta}r^4 N 
\left[
1+\frac{N\delta'}{r}
-N\left(\frac{N'}{2N}-\delta'\right)^2
+\frac{1}{r^4}
\left(
\frac{Q^2}{f(\phi )}+P^2 f(\phi )
\right)
\right]=u_0 \ ,
\end{eqnarray}
where the constant $u_0$ is fixed by the asymptotics. 

To assess possible singular behaviours we remark that the expression of the Ricci and  Kretschmann scalars 
for the line-element (\ref{metric}) read:
\begin{eqnarray}
\label{Ricci}
 &&
R=\frac{N'}{r}(3r\delta'-4)+\frac{2}{r^2}
\left\{
1+N\left[r^2\delta''-(1-r\delta')^2\right]
\right\}
-N'' \ ,
\\
\nonumber
\label{Kr}
&&
K=\frac{4}{r^4}(1-N)^2
+\frac{2}{r^2}
\left[
N'^2+(N'-2N\delta')^2
\right]
+
\left[
N''-3\delta'N'+2N(\delta'^2-\delta'')
\right]^2 \ .
\end{eqnarray}

\subsection{Asymptotic forms of the solutions}
To construct BH solutions, we assume the existence of a horizon located
at $r=r_H>0$.
In its exterior  neighbourhood, one finds the following 
approximate solution, valid for non-extremal BHs:
\begin{eqnarray}
\label{horizon1}
&&
m(r)=\frac{r_H}{2}+m_1(r-r_H)+\dots\ , \qquad
\delta(r)=\delta_0 + \delta_1 (r-r_H)+\dots\ ,
\\
&&
\nonumber
\phi(r)=\phi_0 + \phi_1 (r-r_H)+\dots \ , \qquad 
V(r)=v_1 (r-r_H)+\dots \ ,
\end{eqnarray}
where out of the six parameters, $m_1,\delta_0,\delta_1,\phi_0,\phi_1,v_1$, only two are essential,
$\phi_0$
and 
$\delta_0$, the remaining being determined in terms of these and the global charges as:
\begin{eqnarray}
\label{horizon2}
&&
m_1=\frac{1}{2r_H^2} \left[\frac{Q^2}{f(\phi_0)} + f(\phi_0)P^2 \right] \ , \qquad 
 \phi_1=\frac{df(\phi)}{d\phi}\bigg |_{\phi_0}\frac{1}{2r_H} 
\frac{\left[\frac{Q^2}{f(\phi_0)} - f(\phi_0)P^2 \right]}{\left[\frac{Q^2}{f(\phi_0)} + f(\phi_0)P^2 -r_H^2 \right]} \ , 
\nonumber
\\
&&
\delta_1=-r_H \phi_1^2\ , \qquad
 v_1=\frac{e^{-\delta_0}Q}{r_H^2 f(\phi_0)} \ .
\end{eqnarray}
Note that a similar result holds when considering higher order terms in
the approximate solution (\ref{horizon1}).

For large $r$, one finds the following asymptotic expansions:
\begin{equation}
\label{inf1}
m(r)=M-\frac{Q^2+P^2+Q_s^2}{2r}+\dots,~\phi(r)=\frac{Q_s}{r}+\dots,~
V(r)=\Phi_e-\frac{Q}{r}+\dots,~
\delta(r)=\frac{Q_s^2}{2r^2}+\dots.~{~~}
\end{equation}
The essential parameters introduced in the expansion at  infinity 
(\ref{inf1})
are the ADM mass $M$,
electric and magnetic charges $Q,P$, electrostatic potential at infinity $\Phi_e$
and scalar 'charge' $Q_s$.

The Ricci scalar (\ref{Ricci}) 
vanishes as $r\to r_H$,
while the Kretschmann scalar 
(\ref{Kr}) 
reads
\begin{eqnarray}
\label{Kr-horizon2}
K=\frac{12}{r_H^4}
\left\{
1-\frac{2}{r_H^2}\left[\frac{Q^2}{f(\phi)}+f(\phi)P^2\right]+\frac{5}{3 r_H^4}\left[\frac{Q^2}{f(\phi)}+f(\phi)P^2\right]^2
\right\}+\mathcal{O}(r-r_H) \ .
\end{eqnarray}

\subsection{Quantities of interest and Smarr law}
Two  horizon physical quantities of interest are the Hawking temperature and horizon area
\begin{eqnarray}
T_H=\frac{1}{4\pi}N'(r_H)e^{-\delta_0} \ , \qquad A_H=4\pi r_H^2 \ ;
\end{eqnarray}
these, together with the horizon scalar field value $\phi_0$ compose the relevant horizon data.

The Smarr-like relation~\cite{Smarr:1972kt} for this family of models turns 
out to  have no \textit{explicit} imprint of the scalar hair,
\begin{eqnarray}
\label{Smarr}
M=\frac{1}{2} T_H A_H+\Phi_e Q +\Phi_m P \ ,
\end{eqnarray}
where we have defined a `magnetic' potential as
$\Phi_m\equiv \int_{r_H}^{\infty} dr  e^{-\delta} f(\phi)P/r^2$.
One can then compute a first law of BH thermodynamics for EMS BHs, that reads:
\begin{eqnarray}
\label{1st}
dM=\frac{1}{4}T_H dA_H+\Phi_e dQ +\Phi_m dP \ .
\end{eqnarray}

A non-linear Smarr relation ($i.e.$ mass formula) can also be established for this family of models,
\begin{eqnarray}
\label{figen}
 M^2+Q_s^2=Q^2+P^2+\frac{1}{4}A_H^2 T_H^2 \ ,
\end{eqnarray}
which is derived 
by evaluating the expression of the first integral 
(\ref{Eqs-FistInt})
at the horizon and at infinity,
for the approximate form of the solutions 
(\ref{horizon1})
 and (\ref{inf1}), respectively.

Finally, one can prove that the solutions satisfy the
  virial identity, which is obtained by a Derrick-type~\cite{Derrick:1964ww} scaling argument, see $e.g.$~\cite{Herdeiro:2015waa}
\begin{equation}
\label{virial} 
\int_{r_H}^\infty dr \left\{
e^{-\delta}  \phi'^2
  \left[
1+\frac{2r_H}{r}\left(\frac{m}{r}-1\right) 
   \right]
\right\}
=
\int_{r_H}^\infty dr \left\{
e^{-\delta }  
\left(1-\frac{2r_H}{r}\right)\frac{1}{r^2}
\left[
\frac{Q^2}{f(\phi)}+f(\phi)P^2
\right]
\right\} \ .
\end{equation}
One can show that $1+\frac{2r_H}{r}(\frac{m}{r}-1) >0$,
$i.e.$ the left hand side integrand, is strictly positive. Thus, the virial identity shows that  a nontrivial scalar field requires a nonzero electric/magnetic charge so that the right hand side is nonzero.

The model possesses the scaling
symmetry
 \begin{eqnarray}
\label{scale} 
 r \to \lambda r \ , \qquad (P,Q)\to \lambda(P,Q) \ ,
\end{eqnarray}
where $\lambda>0$ is a constant. Under this scaling symmetry, all other quantities change accordingly, $e.g.$, $M \to \lambda M$,
while 
the coupling function $f(\phi)$
is unchanged.
Thus, for a physical discussion, we consider quantities which are invariant under the transformation (\ref{scale}).
Consequently, we introduce the standard reduced quantities
\begin{eqnarray}
\label{scale1}
q\equiv \frac{\sqrt{Q^2+P^2}}{M}\ , \qquad a_H\equiv \frac{A_H}{16\pi M^2}\ , \qquad t_H\equiv 8\pi T_H M \  .
\end{eqnarray}
For example, dyonic RN BHs have closed expressions for $a_H,t_H$:
\begin{eqnarray}
\label{RNscale}
a_H^{(\rm RN)}=\ \frac{1}{4}(1+\sqrt{1-q^2})^2\ , \qquad t_H^{(\rm RN)}=\frac{4\sqrt{1-q^2}}{(1+\sqrt{1-q^2})^2} \ .
\end{eqnarray}
In Appendix A we exhibit the corresponding expressions for other dilatonic BHs known in closed analytic form, which are class I solutions.

The generic dilatonic  dyonic solutions are not known in closed form,
which hold also for all scalarised BHs.
These solutions are found numerically, by matching the 
asymptotics 
(\ref{horizon1}),
(\ref{inf1}).
Equations 
(\ref{ec-m})-(\ref{ec-phi})
are solved by using a standard Runge-Kutta ODE solver and implementing a shooting method in terms of the parameters
$\phi_0$, $\delta_0$.

\subsection{The BH solutions}

\subsubsection{The purely electric BHs}

\begin{figure}[ht!]
\begin{picture}(0,0)
\put(72,-8){{\bf Class I} (dilatonic)}
\put(288,-8){{\bf Class IIA} (scalarised)}
\end{picture}
\begin{center}
\includegraphics[height=.34\textwidth, angle =0 ]{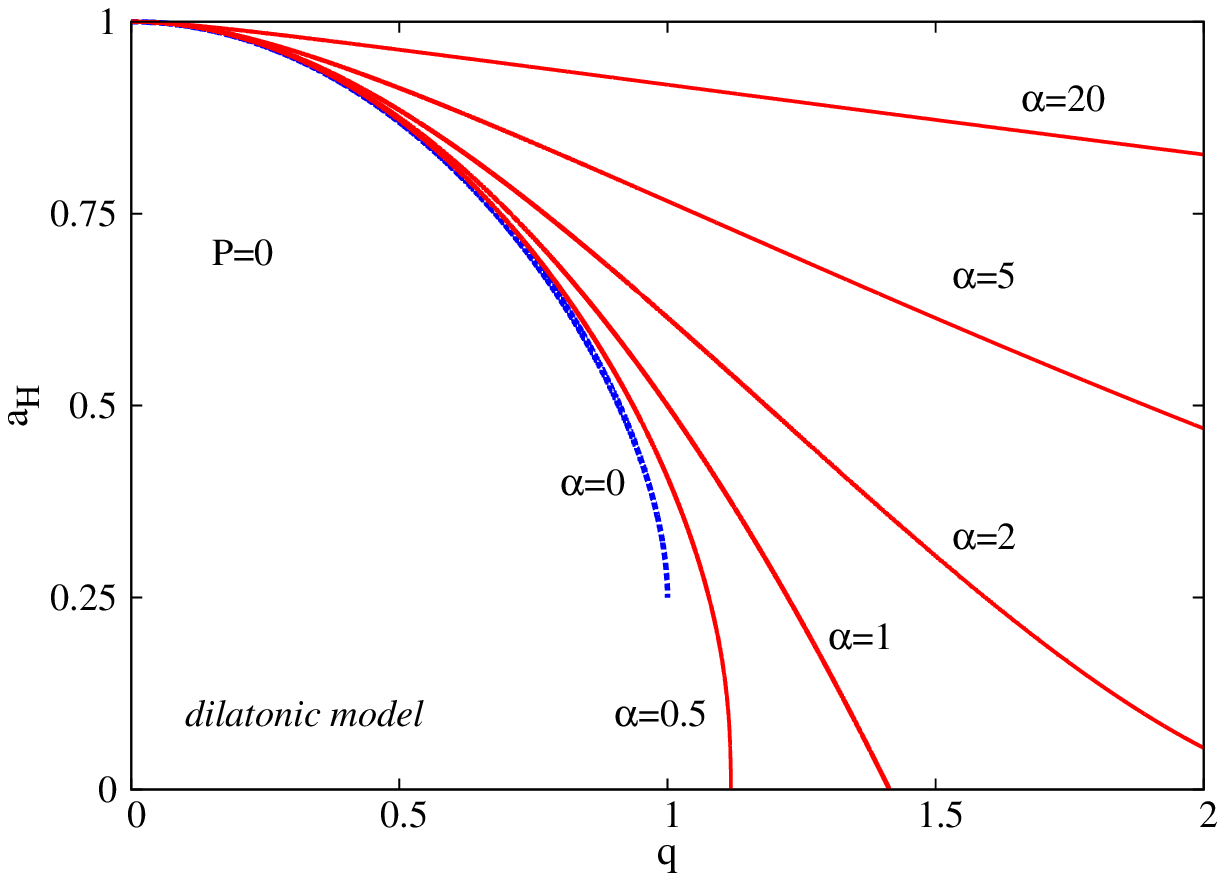} 
\includegraphics[height=.34\textwidth, angle =0 ]{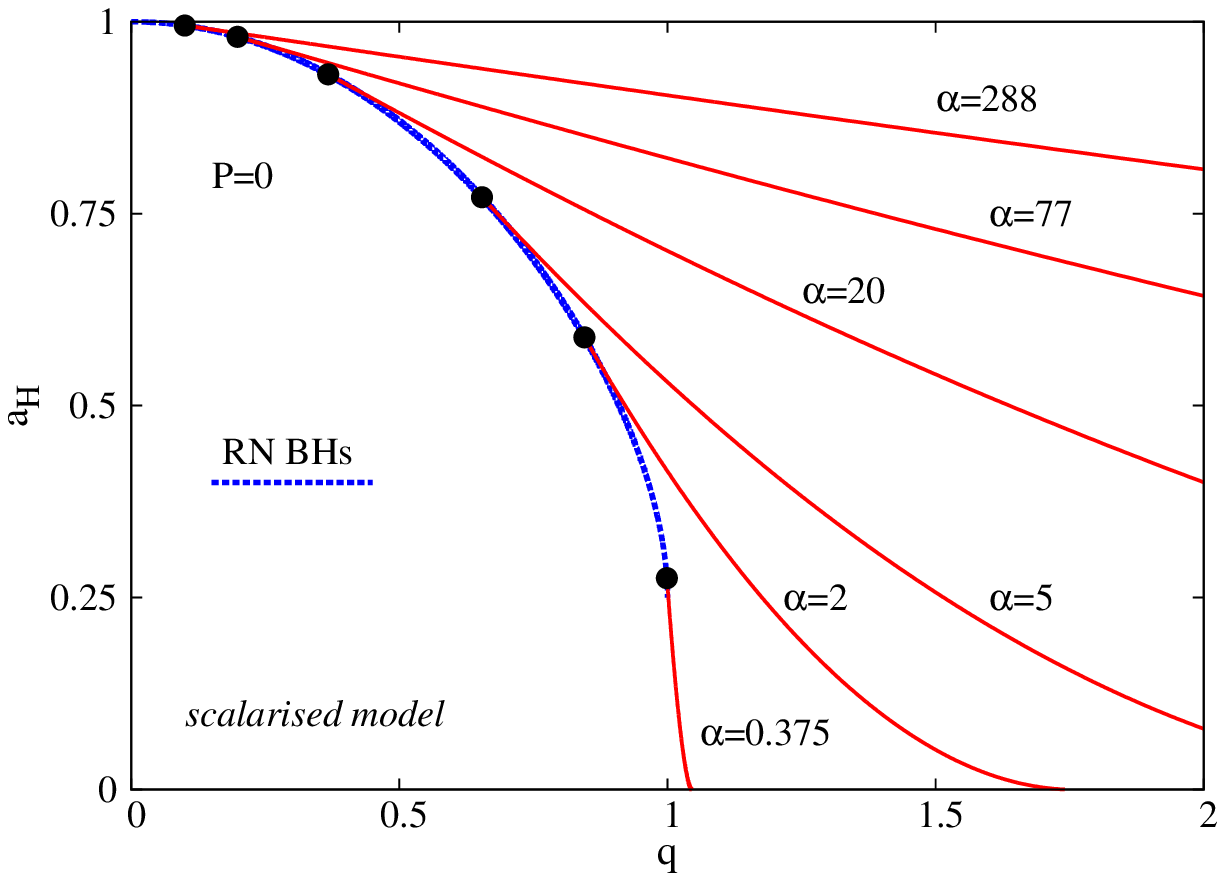}
\includegraphics[height=.34\textwidth, angle =0 ]{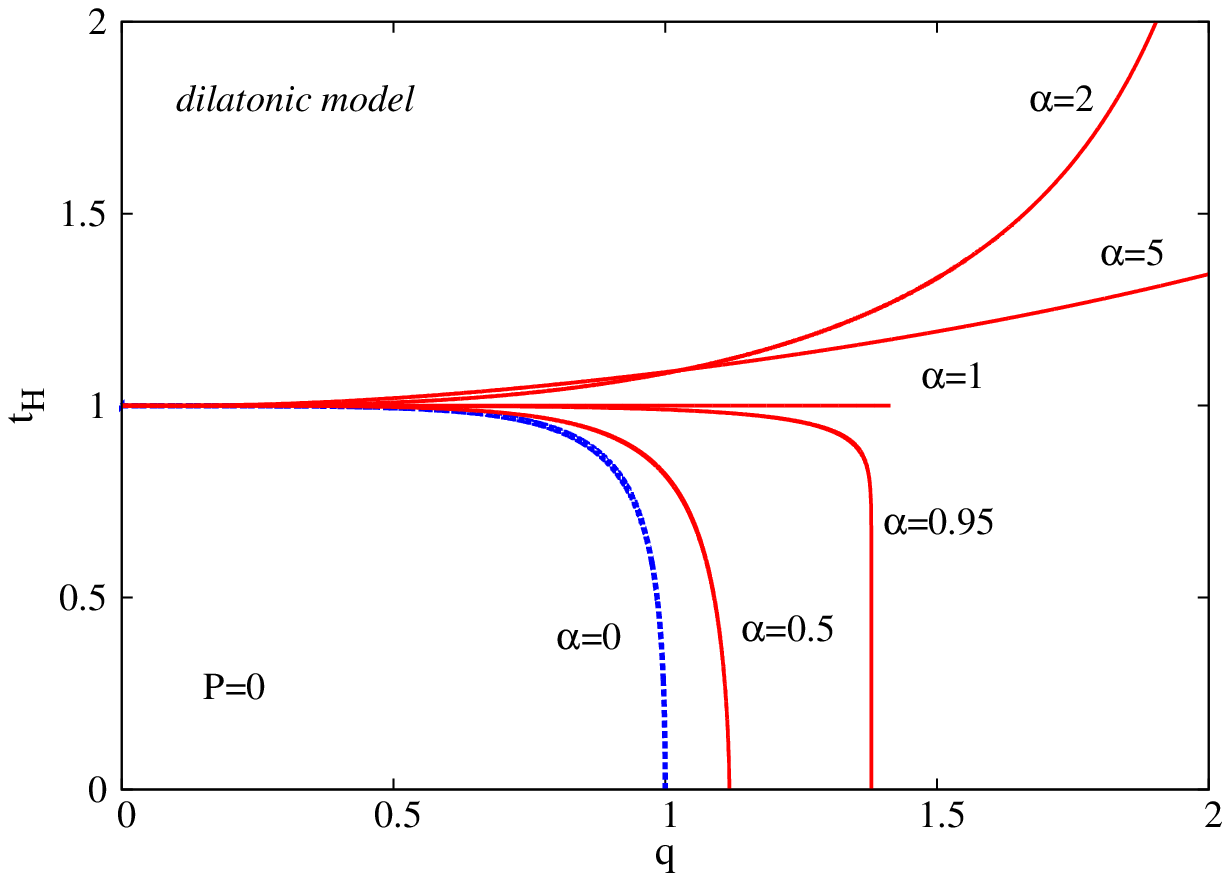} 
\includegraphics[height=.34\textwidth, angle =0 ]{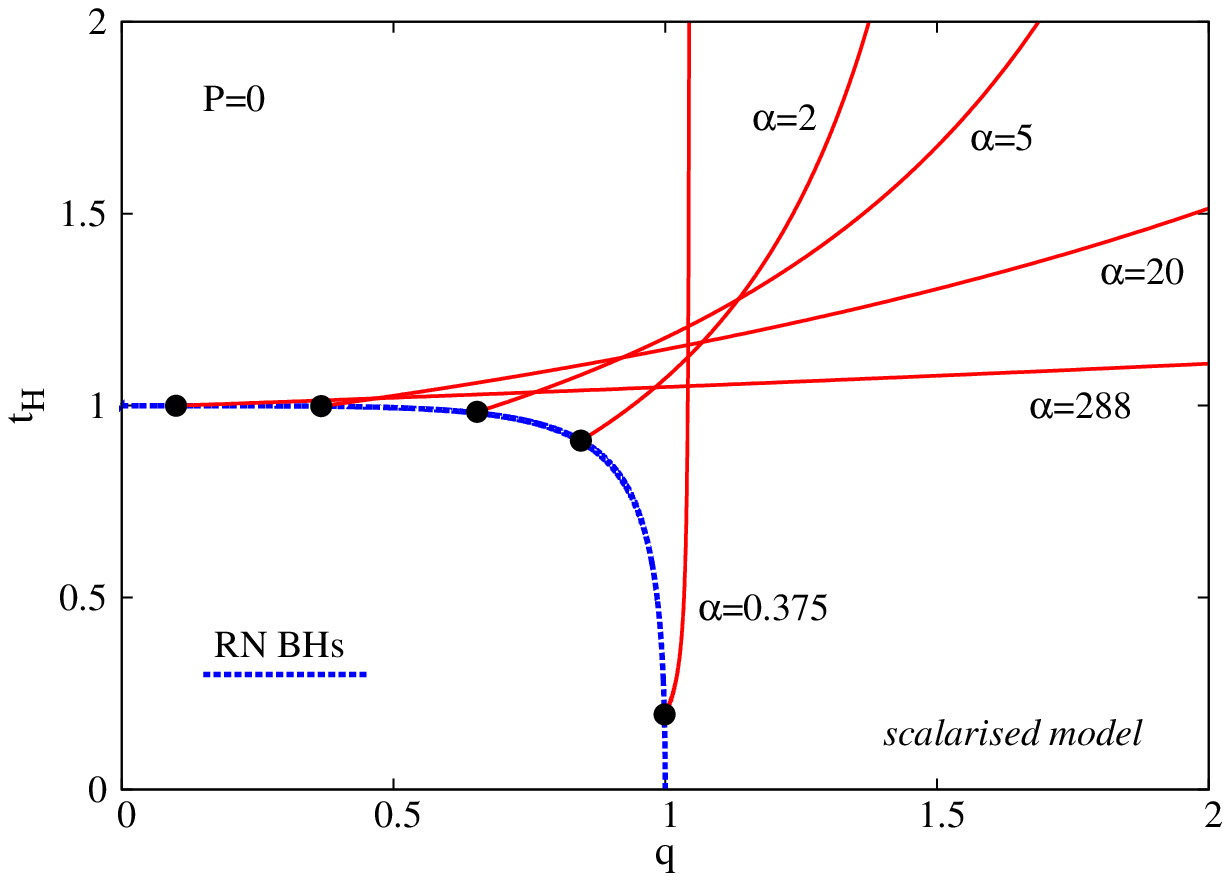}
\end{center}
\caption{
Reduced area $a_H$ (top panels) and reduced temperature $t_H$ (bottom panels) $vs.$ reduced charge
$q$ for dilatonic (left panels) and scalarised solutions  (right panels). 
All solutions have $P=0$.
The blue lines are the set of RN BHs ($\phi=0$).
	 The red lines are sequences of BHs with a nontrivial scalar field for a given $\alpha$. 
	Different sequences are presented, for a range of values of $\alpha\ze $. The black dots indicate
	the RN solutions from which the scalarised BHs bifurcate.
}
\label{electric}
\end{figure}

Let us start with our reference class I solutions. The behaviour of the dilatonic BHs with any $\alpha>0$ is rather similar, albeit  $\alpha=1$ is a somewhat special point that separates the family into two subsets with respect to the behaviour of some physical quantities. This can be seen from the study of the exact solutions in Appendix A.
For any given $\alpha$,
the branch of dilatonic BHs bifurcates from the Schwarzschild BH ($q=0$), rather than RN BHs, 
and ends in a critical solution which is approached for a certain maximal $q$
\begin{eqnarray}
q_{\rm max}^{(\rm D)}=\sqrt{1+\alpha^2} \ ,
\end{eqnarray}
where the superscript `D' refers to dilatonic. The critical solution has, for any $\alpha>0$, a singular horizon, as one can see by evaluating 
the expression (\ref{Kr-horizon2}). The reduced temperature $t_H$, on the other hand, 
goes to zero for $\alpha<1$ and diverges 
for $\alpha>1$. The solutions with $\alpha=1$ have $t_H=1$.
These features can be seen in Fig.~\ref{electric} (left panels),
where the behaviour of $a_H,t_H$ $vs.$ $q$ are illustrated for several values of $\alpha$.

Let us now turn to our reference class IIA solutions. For the purely electric case these have been constructed in~\cite{Herdeiro:2018wub,Fernandes:2019rez}. Let us briefly review their basic properties, emphasizing a comparison with class I solutions.
Given a value of the coupling constant $\alpha>1/8$, 
the spherically symmetric scalarised BHs bifurcate from the corresponding RN BH, 
with a given $q=q(\alpha)\neq 0$, as discussed above.
Keeping constant the parameter $\alpha$, this branch
has a finite extent,
ending again in a critical configuration.
This limiting solution possess a singular horizon,
as found when evaluating the Kretschmann scalar (\ref{Kr-horizon2}).
The horizon area tends to zero as the critical solution is approached and the temperature diverges, 
while the mass and scalar charge remain finite. This behaviour parallels that of the dilatonic solutions with $\alpha>1$.
In the region of the parameter space wherein scalarised and RN BHs exist for the same $q$,
one always finds that the scalarised solutions are entropically favoured
over the RN BHs, as it is manifest from the top right panel of Fig.~\ref{electric}.

The domain of existence of the purely electrically charged BHs of both types will be exhibited below in Fig.~\ref{dyon-domain}, where we also compare it with the dyonic case that we shall discuss next. This domain of existence, in an $(\alpha,q)$-diagram is bounded by two curves. In the dilatonic case, these curves correspond to the Schwarzschild BH and the line of critical solutions. In the scalarised case, these curves correspond to the aforementioned existence line - the set of RN solutions from which the scalarised BHs bifurcate - and, again, the line of critical solutions.

Finally, we remark that, for both dilatonic and scalarised solutions,
along any branch with fixed  $\alpha$,
the ratio $q={\sqrt{Q^2+P^2}}/M$
increases and becomes larger than unity at some stage. In this sense, \textit{overcharged} BHs are possible, in contrast with the RN family.

\subsubsection{The dyonic BHs}
\begin{figure}[ht!]
\begin{center} 
\includegraphics[height=.34\textwidth, angle =0 ]{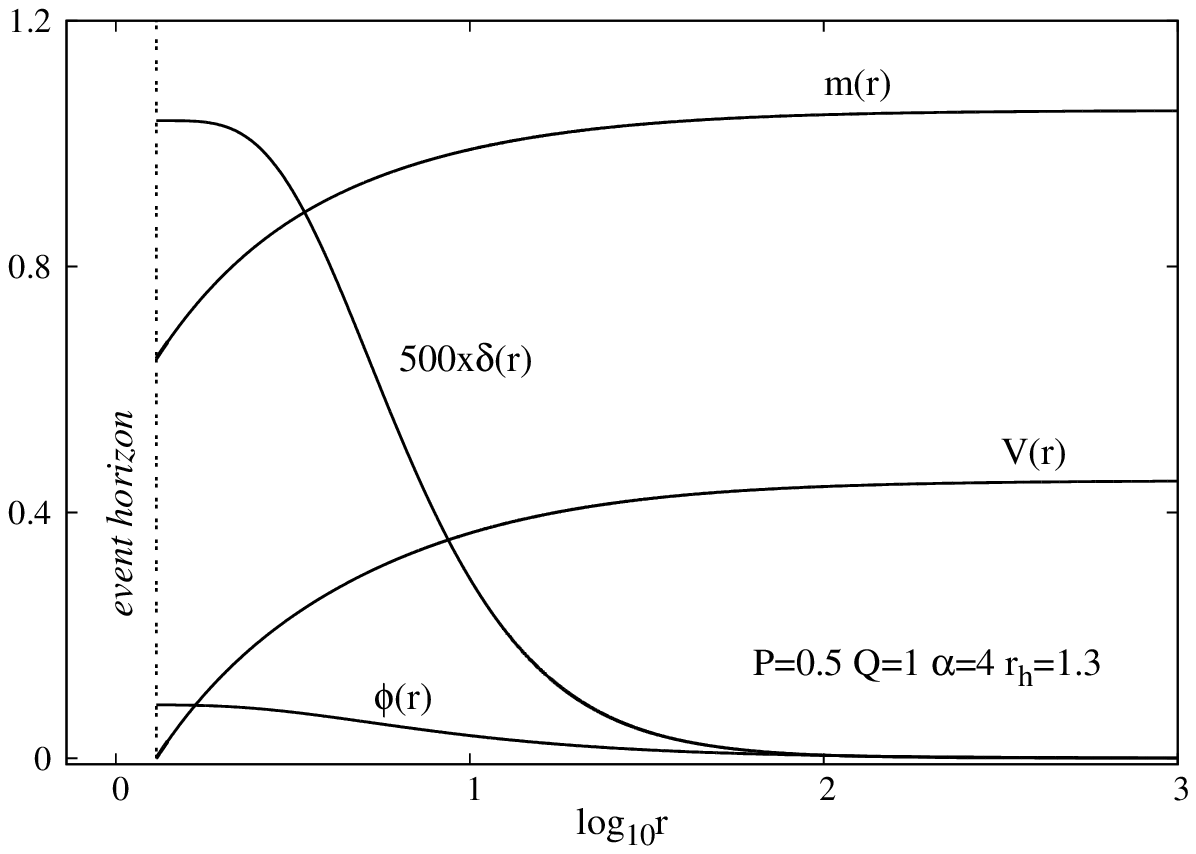}
\includegraphics[height=.34\textwidth, angle =0 ]{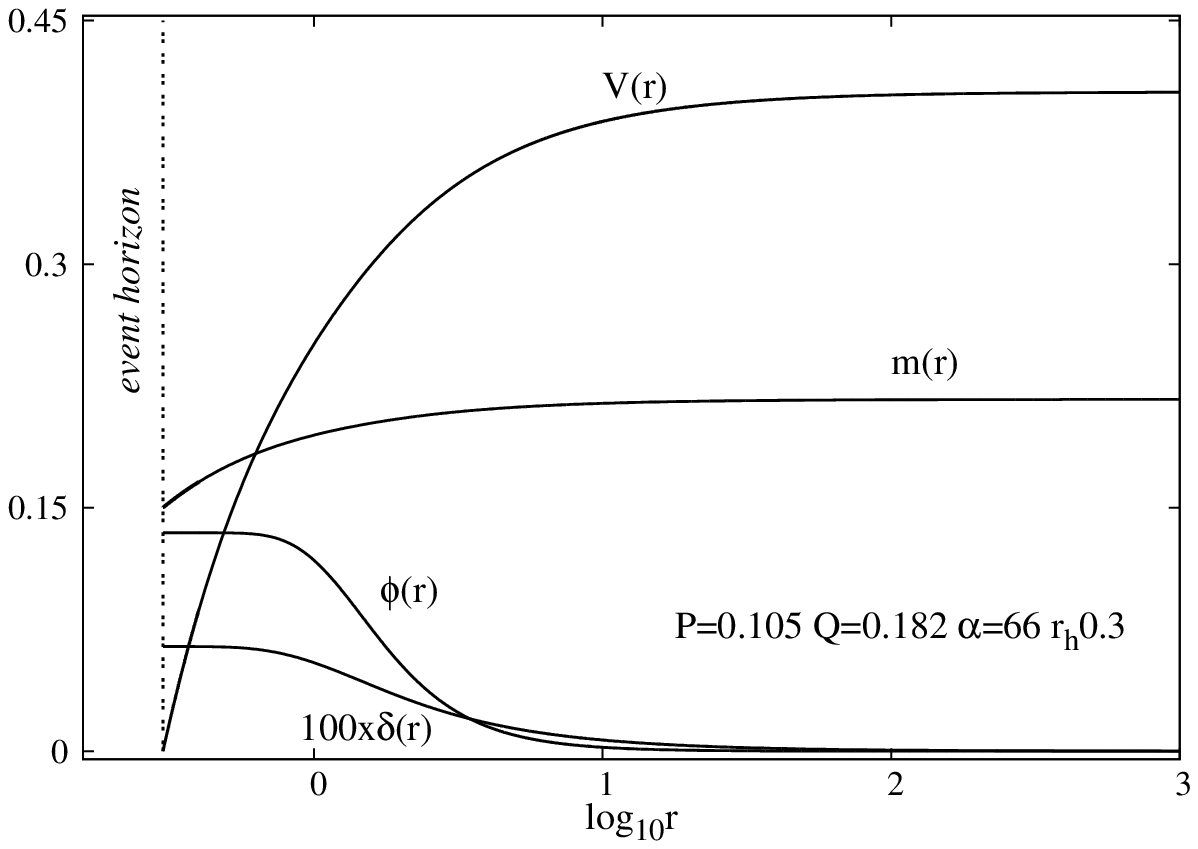}
\end{center}
\caption{
Examples of dyonic BHs radial profile functions for a dilatonic (left panel) and a scalarised (right panel) BH.}
\label{dyon-profile}
\end{figure}

The purely electric solutions above, for both classes discussed, possess generalisations with a nonzero magnetic charge.
The profile functions of  illustrative dyonic BHs are shown in Fig.~\ref{dyon-profile}, 
for both the dilatonic and scalarised cases.

\begin{figure}[ht!]
\begin{picture}(0,0)
\put(72,-8){{\bf Class I} (dilatonic)}
\put(288,-8){{\bf Class IIA} (scalarised)}
\end{picture}
\begin{center}
\includegraphics[height=.34\textwidth, angle =0 ]{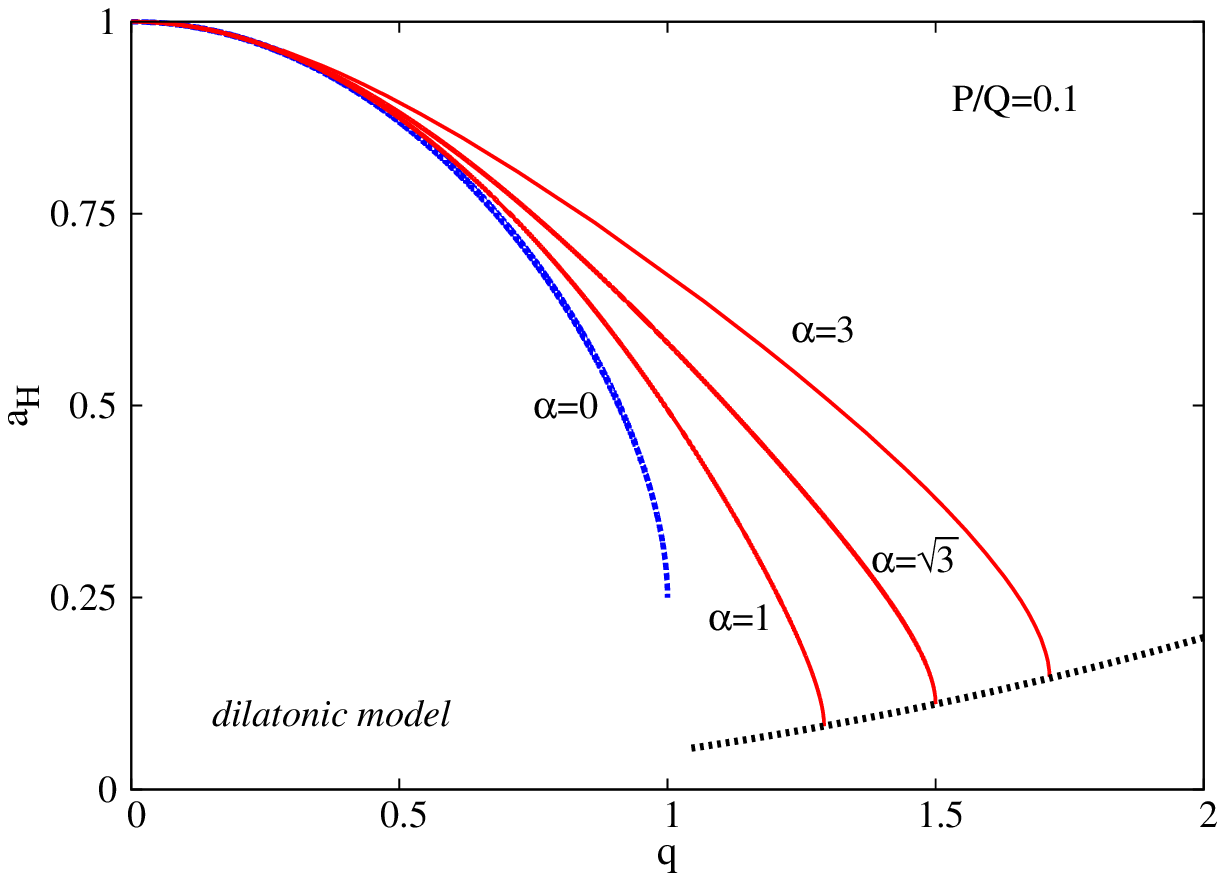} 
\includegraphics[height=.34\textwidth, angle =0 ]{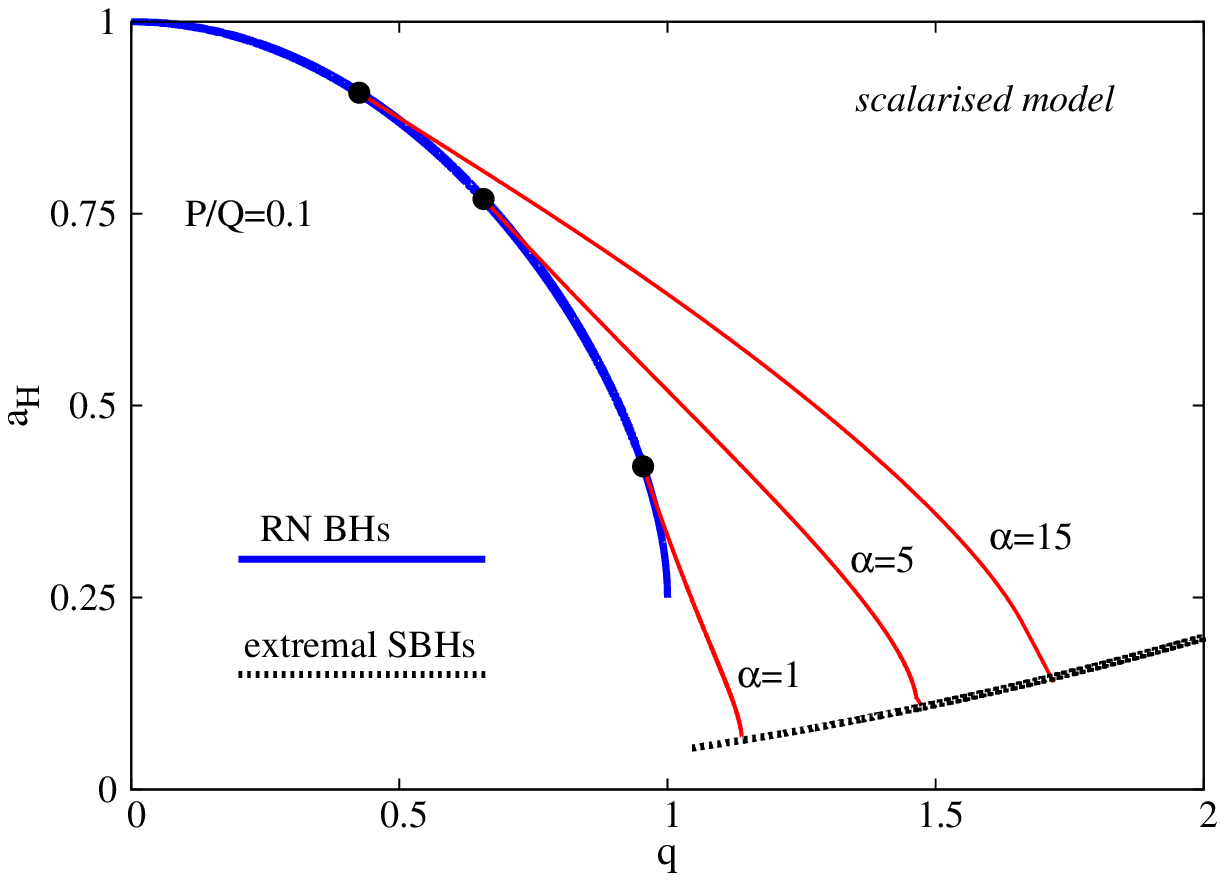}
\includegraphics[height=.34\textwidth, angle =0 ]{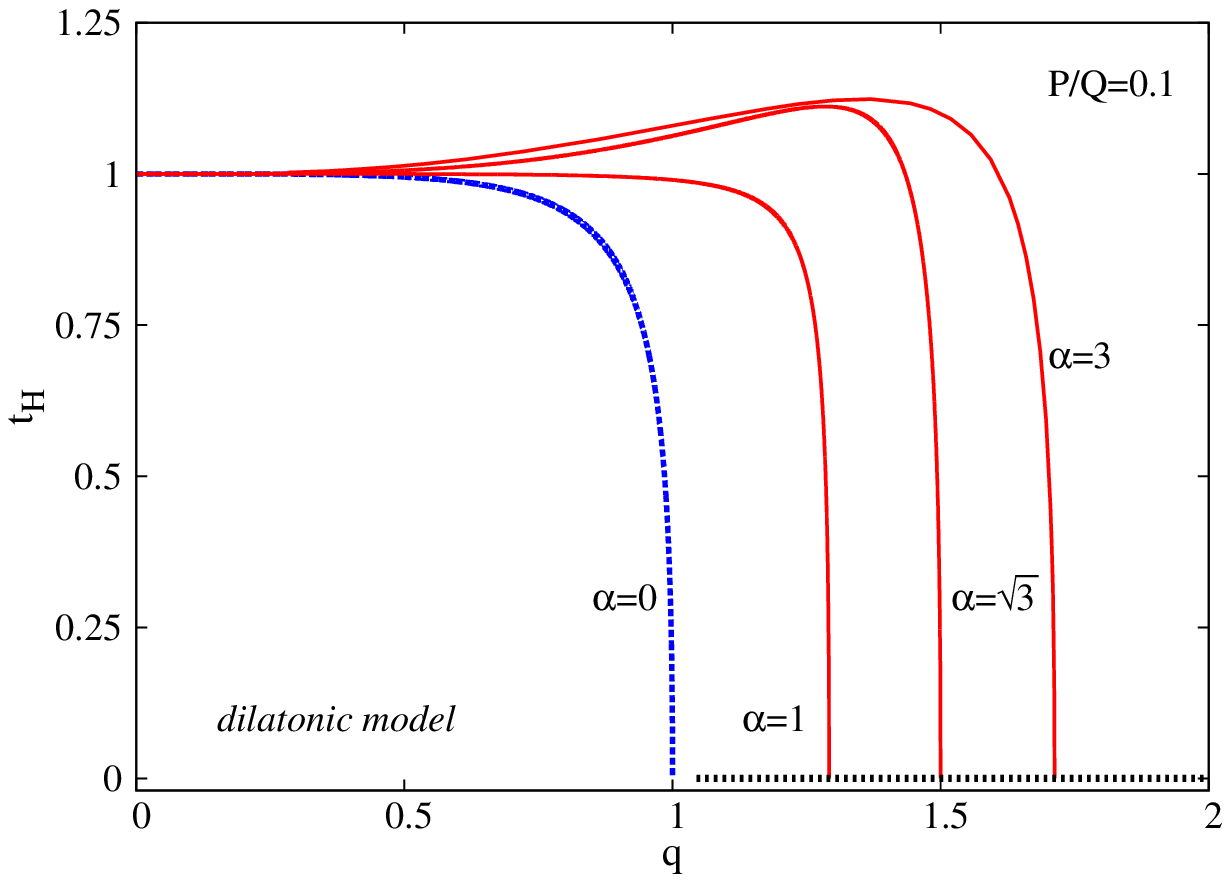} 
\includegraphics[height=.34\textwidth, angle =0 ]{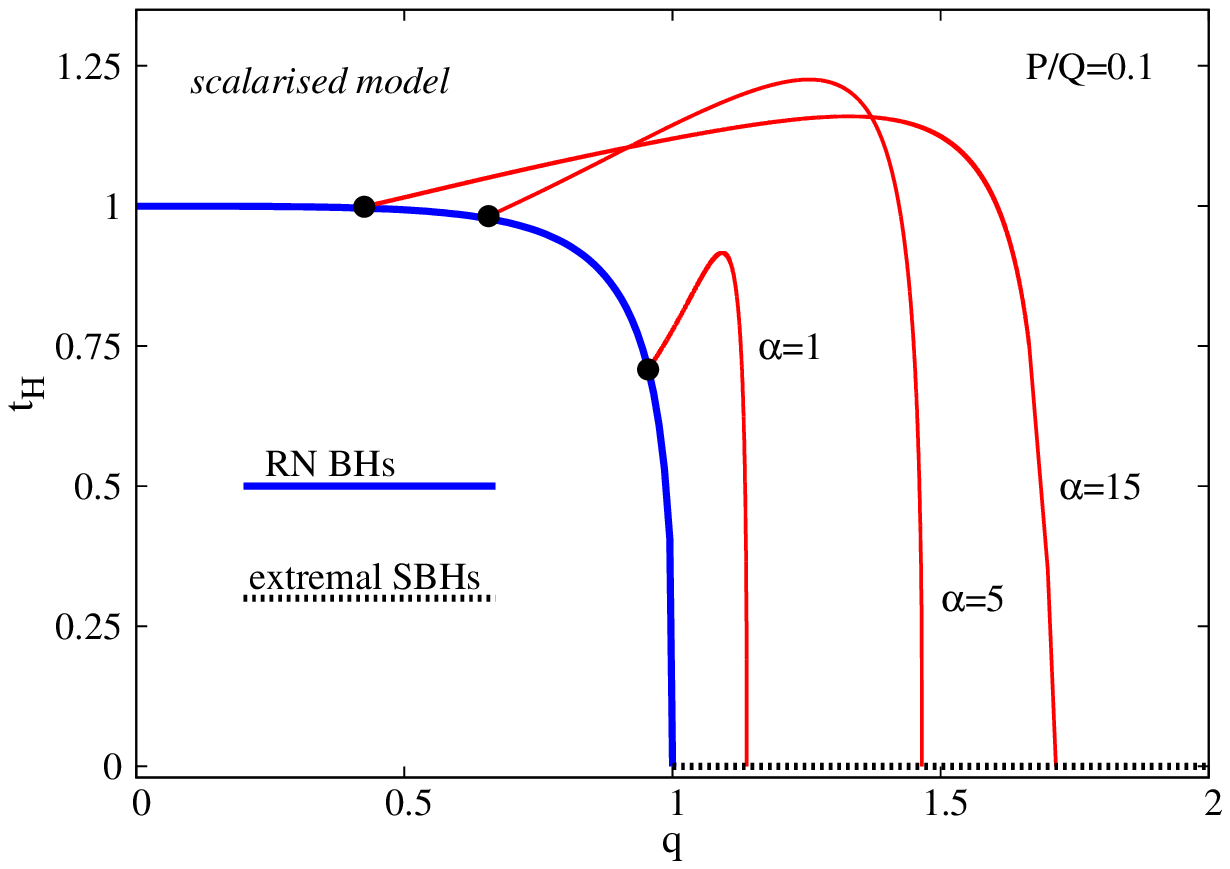}
\end{center}
\caption{
Same as Fig. 1 but now  for dyonic BHs. The ratio between magnetic and electric charges
is $P/Q=0.1$.
}
\label{dyons}
\end{figure}

Dyonic BHs preserve some, but not all, of the qualitative characteristics 
of the purely electric solutions.
In the dilatonic case, the
branch of solutions with a given $\alpha$
 starts again from the Schwarzschild limiting solution 
(which has $a_H=1$, $t_H=1$ and $q=0$)
and ends in a limiting configuration with 
$a_H>0$, $t_H=0$ and $q=q_{\rm max}>0$ - Fig.~\ref{dyons} (left panels).
This limiting solution, however, is now an 
extremal BH (rather than a singular critical solution) and will be discussed in the next Section.	

Unlike the dilatonic solutions, which exist for arbitrarily small $q$ for any $\alpha$,  scalarised BHs 
 with a given $\alpha$
exist for $q>q_{min}$ only.
They
bifurcate from a RN BH (with $q>0$)
and form a branch ending again on an extremal solution 
with vanishing horizon temperature and nonzero horizon area - Fig.~\ref{dyons} (right panels).
As for purely electric solutions,  for the same global charges $M,P,Q$,  
the scalarised solutions are entropically preferred over the corresponding RN solution.

\begin{figure}[ht!]
\begin{center}
\includegraphics[height=.34\textwidth, angle =0 ]{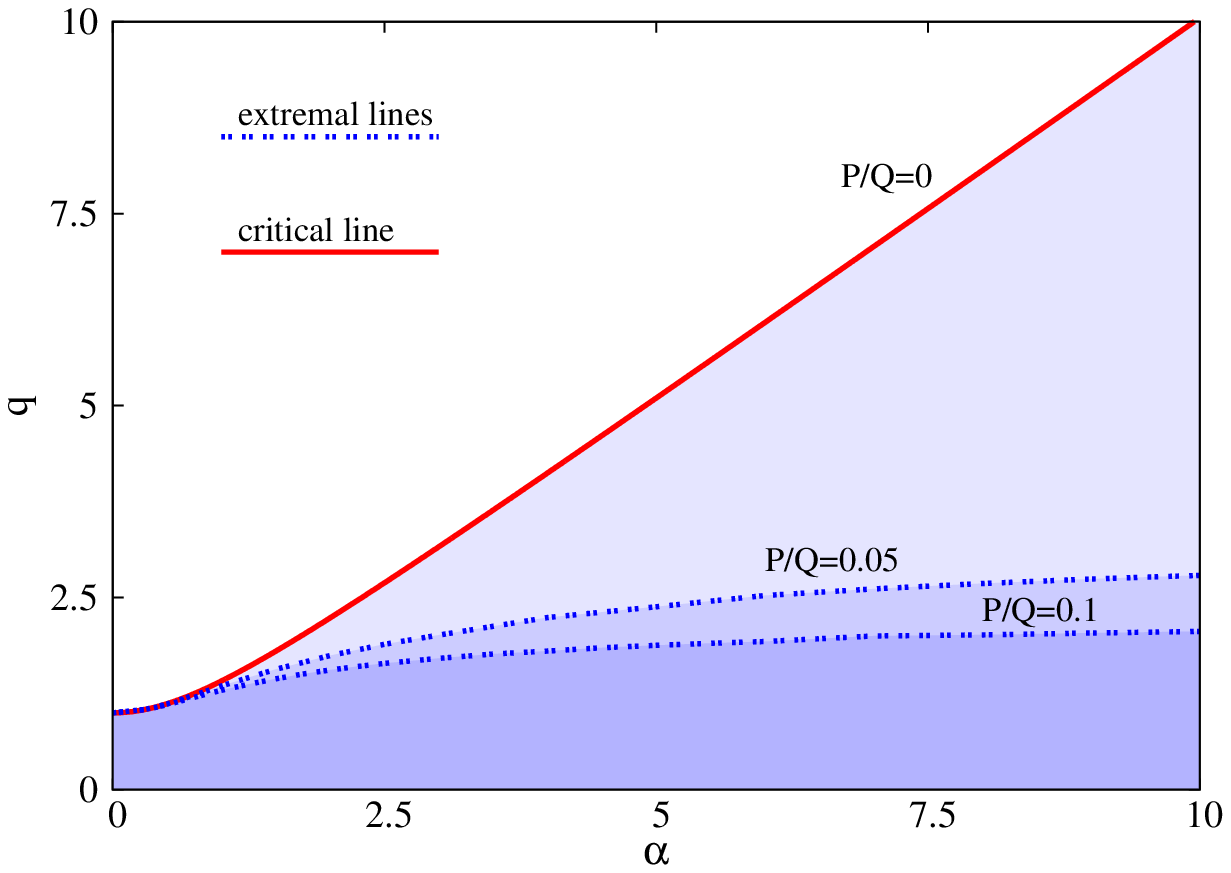}
\includegraphics[height=.34\textwidth, angle =0 ]{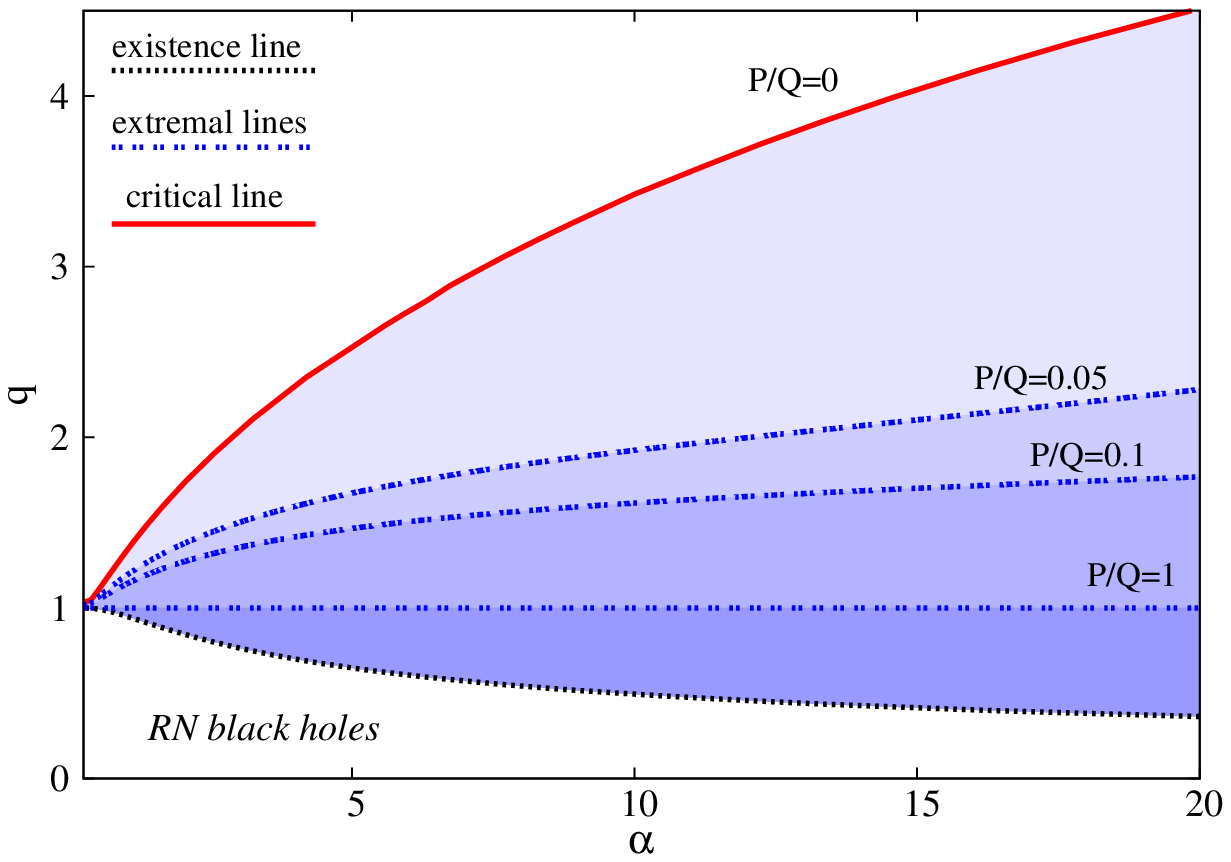}
\end{center}
	 	 \caption{Domain of existence of dilatonic BHs (left panel) and scalarised BHs (right panel)
		for several values of the ratio $P/Q$. 
	}
\label{dyon-domain}
\end{figure}

The domain of existence of the dyonic BHs is shown in Fig.~\ref{dyon-domain}
for several values of the ratio $P/Q$ and for both dilatonic and scalarised BHs.
In particular, observe that in both cases,
the maximal allowed value of $q$  for BHs with
a given $\alpha$ decreases as the ratio $P/Q$ increases. In other words, the domain of existence shrinks, as the magnetic charge is increased, for fixed $Q$.

\section{Extremal BHs}
\label{sec4}

\subsection{Numerical construction}
To address extremal BHs one needs to impose a different near-horizon expansion to that in~\eqref{horizon1}, which accounts for the degenerate  horizon. 
The leading order terms of the appropriate expansion are:
\begin{eqnarray}
\label{extr1}
&&
N(r)=N_2(r-r_H)^2+\dots\ , \qquad
\delta(r)=\delta_0+\delta_1 (r-r_H)^{2k-1}+\dots \ , \nonumber
\\
&&
\label{extr12}
\phi(r)=\phi_0+\phi_c (r-r_H)^{k}+\dots \ , \qquad
V(r)= v_1(r-r_H)+\dots \ .
\end{eqnarray}
One can show that the next to leading order term  in the expression of $N(r)$ is $\mathcal{O}(r-r_H)^3$.
It is convenient to take 
 $r_H$ and $\phi_0$
as essential parameters.
Then the field equations imply
\begin{eqnarray}
\label{extr2}
 Q =\frac{r_H\sqrt{f(\phi_0)}}{\sqrt{2}}\ , \qquad  P=\frac{r_H}{\sqrt{2f(\phi_0)}} \ , \qquad N_2=\frac{1}{r_H^2} \ .
\end{eqnarray}
Consequently, given an expression of the coupling function 
$f(\phi)$,
 one can  express the value of the scalar field at the horizon
$\phi_0$ as a function of $P,Q$,  
by solving the equation
\begin{eqnarray}
\label{extr21}
f(\phi_0)=\frac{Q}{P}\ , \qquad {\rm while}~~r_H=\sqrt{2 PQ}\ .
\end{eqnarray}
The expansion  (\ref{extr1})
contains two free parameter $\phi_c$ and $\delta_0$
which are fixed by numerics, while $\delta_1,v_1$ are fixed as 
\begin{eqnarray}
\label{extr4}
\delta_1=-\frac{r_H\phi_c^2k^2}{2k-1}\ , \qquad  v_1=\frac{e^{-\delta_0}Q}{r_H^2 f(\phi_0)} \ .
\end{eqnarray}

The power $k$
in (\ref{extr1}) is given by
\begin{eqnarray}
\label{extr3}
k=\frac{1}{2}\left(-1+\sqrt{1+2 \left(\frac{f'(\phi_0)}{f(\phi_0)} \right)^2} \right)>0 \ ,
\end{eqnarray}
which, generically, takes non-integer values.
However, 
a non-integer $k$ implies that a sufficiently higher order derivative of the curvature tensor will diverge as $r\to r_H$.
A minimal requirement for smoothness is that the metric functions $N,\delta$  and their first and second derivatives are finite as $r\to r_H$; this yields the condition
 \begin{eqnarray}
 \label{constr1}
  k >3/2 \ .
 \end{eqnarray} 
On the other hand, for analytic solutions on the horizon (as extremal RN), the power $k$ in the above 
near horizon
expansion (\ref{extr1}), (\ref{extr12})
should be an integer. This imposes the condition
\begin{eqnarray}
\label{extr5}
\frac{f'(\phi_0)}{f(\phi_0)}= \pm \sqrt{2p(p +1)}\ , \qquad {\rm with}~~p=1,2,\dots \ .
\end{eqnarray}
For the dilatonic case, condition~\eqref{extr5} translates to 
\cite{Galtsov:2014wxl,Zadora:2017qeo}
(see also \cite{Abishev:2015pqa})
\begin{eqnarray}
\label{cond12}
 \alpha=\sqrt{\frac{p(p+1)}{2}} \ ,
\end{eqnarray}
again with an integer $p$.
For scalarised solutions with the coupling function (\ref{quadratic})
the condition (\ref{extr5})
becomes
\begin{eqnarray}
\label{cond2}
 \alpha= \frac{p(p+1)}{4 \log(\frac{Q}{P})}\ .
\end{eqnarray}

The extremal solutions share the far field asymptotics 
(\ref{inf1}) with 
 the non-extremal ones; moreover, the relations
 (\ref{Smarr})-(\ref{figen}) hold also for $T_H=0$.

\begin{figure}[ht!]
\begin{center}
\includegraphics[height=.45\textwidth, angle =0 ]{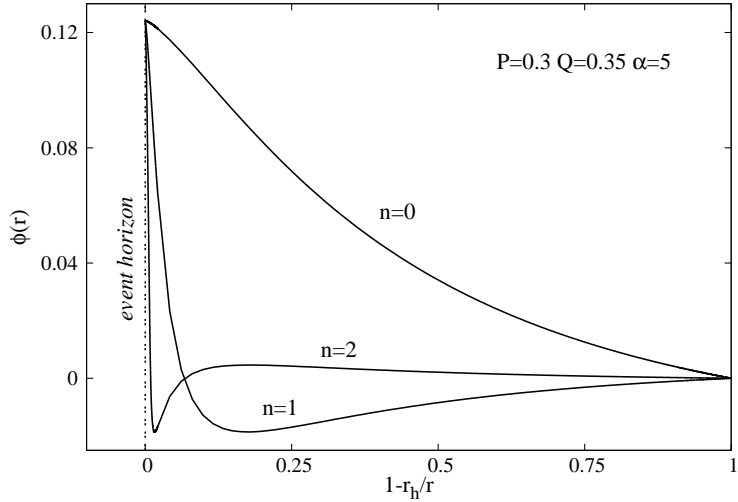}
\end{center}
\caption{
A sequence of scalar field profiles starting with the same horizon data
in a scalarised model. Each solution possesses a different node number.
}
\label{grav2}
\end{figure}

We have constructed extremal solutions  by using the same numerical
approach as in the generic non-extremal case. The profile of the various functions resulting from the integration are not particularly enlightening, resembling those 
in the non-extremal case and shall not be shown here.
But we would like to point out a peculiar feature of the extremal scalarised BHs. There exists a (presumably)
infinite family of solutions with the same horizon data as specified by $(\phi_0,r_H)$ (or, equivalently, $(P,Q)$), labelled by their node-number $n$. This is illustrated in Fig.~\ref{grav2}:  the scalar field always starts at the same horizon value; however,
the bulk profile is different. As expected for excited states, the mass of these solutions increases with $n$. We remark no excited configurations were found in the dilatonic case, which always has $n=0$.

\subsection{An analytic approach: the attractor mechanism and entropy function}
The numerical construction of the extremal BHs is a  difficult numerical task.
Let us now provide a different argument for the existence of the EMS extremal dyonic
BHs:  the existence of an exact solution
describing a Robinson-Bertotti vacuum, namely an $AdS_2\times S^2$ spacetime. 
As for extremal RN BHs, we expect that this solution describes the
neighbourhood of the event horizon of an extremal scalarised BH
with nonzero magnetic and electric charges. As we shall see, both charges are mandatory for the Robinson-Bertotti vacuum to exist with a non-trivial scalar field.

To search for the Robinson-Bertotti vacuum we consider~\eqref{metric-generic} with $a(r)^2=v_0r^2$, $b(r)^2=v_1$, $c(r)^2=v_0/r^2$, that is the line element
\begin{equation}
\label{AdS2S2}
  ds^2=v_0\left(-r^2dt^2+\frac{dr^2}{r^2}\right)+v_1(d\theta^2+\sin^2\theta d\varphi^2) \ ,
\end{equation}
and the matter fields ansatz 
\begin{equation}
\label{matter-attractors}
A=e r dt+P\cos \theta d \varphi\ , \qquad \phi=\phi_0 \ .
\end{equation}
The constant parameters 
$\{v_0,v_1; e,P,\phi_0\}$
satisfy a set of algebraic relations which result from the EMS
equations~(\ref{eqEinstein})-(\ref{eqM}). However, instead of attempting to solve these, we shall, in what follows, determine these parameters
 by using the formalism proposed
in \cite{Sen:2005wa, Astefanesei:2006dd, Sen:2007qy}, 
thus by extremizing an {\it entropy function}.
This approach allows us also to compute the BH entropy
 and to show that the solutions exhibit an attractor-type behaviour.

Let  us consider the Lagrangian density of the model, as read off from (\ref{action}), evaluated for the ansatz (\ref{AdS2S2})-(\ref{matter-attractors}) and integrated over the angular coordinates,
\begin{equation}
  {\cal L} = \frac{1}{16 \pi}\int d\theta d\varphi\sqrt{-g} 
	\left(
	R-2(\nabla \phi)^2 
- f(\phi) F^2  
\right)
=\frac{1}{2}
\left[
 v_0-v_1+f(\phi_0)\left(\frac{e^2 v_1}{v_0}-\frac{P^2 v_0}{v_1}\right)
\right]~.{~~}
\end{equation}
Then, following \cite{Sen:2005wa, Astefanesei:2006dd, Sen:2007qy},
 we define the entropy function ${\cal E}$ 
by taking the Legendre transform
of the above integral with respect to a parameter $Q$,
\begin{eqnarray}
{\cal E}=
  2\pi (e Q -{\cal L})\  ,
\end{eqnarray}
where $Q=\partial {\cal E}/\partial e$ is the electric charge of the solutions.
It follows as a consequence of the equations of motion that the constants 
$\{v_0,v_1; e,\phi_0\}$
 are solutions
of the equations
\begin{equation}
  \frac{\partial {\cal E}}{\partial \phi_0}=0\,, \qquad \frac{\partial {\cal E}}{\partial v_i}=0\,,
  \qquad \frac{\partial {\cal E}}{\partial e }=0\, ,
  \label{attractor}
\end{equation}
or, explicitly,
\begin{eqnarray}
  \label{at1}
  \frac{\partial {\cal E}}{\partial v_0} & = & 0\,\,\,\Rightarrow \,\,\,
	-1+\left(\frac{v_1}{v_0^2}e^2+\frac{1}{v_1} P^2\right)f(\phi_0) =0 \ ,
	\\
  \label{at2}
  \frac{\partial {\cal E}}{\partial v_1} & = & 0\,\,\,\Rightarrow \,\,\,
	1-\left(\frac{1}{v_0} e^2+\frac{v_0}{v_1^2} P^2\right)f(\phi_0)=0 \ ,
	\\
  \label{at3}
  \frac{\partial {\cal E}}{\partial \phi_0} & = & 0\,\,\,\Rightarrow \,\,\,
	\left( P^2 v_0^2-e^2 v_1^2 \right)\frac{d f}{d \phi_0} = 0\ ,
	\\
  \label{at4}
  \frac{\partial {\cal E}}{\partial e} & = & 0\,\,\,\Rightarrow \,\,\, 
	Q = e \frac{v_1}{v_0} f(\phi_0) \ . 
\end{eqnarray} 
The sum of~\eqref{at1} and~\eqref{at2} leads to the generic relation
\begin{eqnarray}
v_0=v_1\ .
\end{eqnarray} 
Thus, the `radius' of the $AdS_2$ is always equal with the one of $S^2$ in the metric (\ref{AdS2S2}). Then, the equation ~\eqref{at4} becomes
\begin{eqnarray}
Q = e  f(\phi_0) \ .
\label{at5}
\end{eqnarray} 
Consequently, eq.~(\ref{at3})
implies the existence of two different families of solutions:
\begin{description}
\item[a)]  eq.~(\ref{at3}) is solved if the coupling function obeys
${d f}/{d \phi_0}=0$. Then, 
 $e$ and $P$ are independent quantities and, from~\eqref{at1},
\begin{eqnarray}
v_0=v_1=(e^2+P^2)f(\phi_0) \ .
\end{eqnarray} 
This family of solutions is only possible in the scalarised case. In this case, ${d f}/{d \phi_0}=0$, with $\phi_0=0$.
Therefore,  one obtains the near horizon geometry of the extremal RN BH, with a vanishing scalar field.
\item[b)]  eq.~(\ref{at3}) is also solved if 
\begin{eqnarray}
\label{cond-a}
 e=P\ , \qquad \stackrel{\eqref{at5}}{\Rightarrow} \qquad Q= P f(\phi_0) \ ,
\end{eqnarray} 
and, from~\eqref{at1},
\begin{eqnarray}
v_0=v_1=2P^2 f(\phi_0) \ .
\end{eqnarray} 
This family of solutions is possible for both the scalarised and dilatonic cases and demands both $Q,P$ to be non-vanishing. 
\end{description}
The scalarisation mechanism is encoded in the existence of two different types of 
attractor solutions in the scalarised EMS models. This contrasts with the case of the dilatonic coupling, for which  condition (\ref{cond-a})
is mandatory and only one type of solutions exists, that requires both electric and magnetic charges to be present.

It is straighforward to check that in both cases the entropy function, 
 ${\cal E}$, evaluated
at the attractor critical point is given by one-quarter of
the area of angular sector in  (\ref{AdS2S2}),
\begin{eqnarray}
S=\pi v_1\ .
\end{eqnarray} 

Finally, we remark that the correspondence of the above parameters with the
ones in the near horizon expansion of the extremal BHs in Section 4.1 is straightforward:
\begin{equation}
v_1=r_H^2 \ , \qquad v_0=1/N_2 \ .
\end{equation}
%

\section{Discussion}
\label{sec5}

In this paper we have investigated the properties of static electromagnetically charged BHs with a non-trivial scalar field profile in EMS models, which are described by~\eqref{action}. A natural classification of these EMS models arises from the standard RN BH of electrovacuum being, or not, a solution. This divides EMS models into two classes. Class I, or dilatonic-type, does not admit RN as a solution. We have illustrated this class by a well-known family of \textit{dilatonic} BHs that naturally emerge in the low energy limit of string theory, as well as in Kaluza-Klein theories. Class II, or scalarised-type, admits RN as a solution. The RN BH may, or may not, be continuously connected to the new BHs with a scalar field profile, naturally leading to two subclasses. In class IIA RN is continuously connected to the new BHs. This class contains the models wherein spontaneous scalarisation of the RN BH occurs~\cite{Herdeiro:2018wub}, dynamically leading to the new \textit{scalarised} BHs. We have illustrated this class by a particular choice of coupling function, introduced in~\cite{Herdeiro:2018wub} in this context. In class IIB, RN is not continously connected to the new BHs and the RN BH is not unstable against scalarisation. 

One of the motivations for this work was to understand the effect of a magnetic charge in the EMS BHs. In the well known dilatonic case, dyonic BHs have a regular extremal limit, whereas purely electrically (or magnetically) charged ones do not; the latter become singular, approaching a critical solution when endowed with the maximal possible charge for a given mass. Given the special features of smooth extremal solutions, it is of interest to understand the status of these solutions in the generic EMS case, since for purely electic scalarised BHs maximal charge led to critical, rather than extremal, solutions~\cite{Herdeiro:2018wub,Fernandes:2019rez}. Here we have shown that for scalarised BHs the conclusion is similar to dilatonic BHs (within a certain coupling regime) in this respect: dyonic BHs can have a regular extremal limit. Our analysis also allows constructing such dyonic extremal solutions for arbitrary coupling in the dilatonic case, since solutions where only known (in analytic closed form) for some particular values of the coupling. Morover, despite the defining difference in the two classes of solutions, Fig.~\ref{electric} and~\ref{dyons} show that these two classes, for the illustrative families, present similar trends in the behaviour of physical quantities.

As evidence for the existence of dyonic extremal scalarised BHs, we have made use of the fact one expects such solutions to have a near-horizon geometry which is, itself, a solution of the field equations. Both for RN and Kerr extremal BHs (when $T_H=0$), the near horizon geometry has an enhanced symmetry that contains an $AdS_2$ geometry 
(for Kerr, there exists a non-trivial fibration of an $S^1$ on $AdS_2$ in the near horizon geometry). 
It was proven in~\cite{Sen:2005wa, Astefanesei:2006dd} that the existence of $AdS_2$ factor is, in fact, at the basis of the attractor mechanism for extremal BHs rather than supersymmetry \cite{Ferrara:1995ih, Ferrara:1996dd}. 
In string theory, the attractor mechanism provides a non-renormalization theorem for the matching of statistical and thermodynamic entropies of extremal BHs \cite{Dabholkar:2006tb} (see, also, section $5$ of \cite{Astefanesei:2006sy}). Here, the attractor mechanism provides a clear 
and simple explanation of why the extremal limit  is a naked singularity for solutions with a single charge and a smooth geometry for dyonic BHs. Besides enabling a partial analytical understanding of the extremal solutions, analysing the near horizon geometry provides an insight on how scalarisation leaves a trace at the level of attractors, allowing two families of near horizon geometries.

Let us close considering some future research. It would be interesting to motivate class II models from a more fundamental viewpoint.  In this respect, we remark that (\ref{action}) may be viewed as a member of a more general family of low energy string theory actions (see, e.g., \cite{Trigiante:2016mnt}).  For example, in four dimensions, the effective string theory can be described by $\mathcal{N}=2$ supergravity (and its deformations \cite{DallAgata:2012mfj}), 
with the generic bosonic Lagrangian density
\begin{equation}
L= 
-\frac{R}{2}
+h_{i\bar{\jmath}}\,\partial_\mu z^i\,\partial^\mu \bar{z}^{\bar{\jmath}}
+\frac{1}{4}\,\mathcal{F}_{\Lambda\Sigma}(z,\bar{z})\,F^\Lambda_{\mu\nu}\,F^{\Sigma\,\mu\nu}
+\frac{1}{8\, e_D}\,\mathcal{R}_{\Lambda\Sigma}(z,\bar{z})\,\eps^{\mu\nu\rho\sigma}\,F^\Lambda_{\mu\nu} \,F^{\Sigma}_{\rho\sigma}
-V(z,\bar{z}) \ .
\label{boslagr}
\end{equation}
The model possesses
 $n_s$ complex scalars $z^i$, $i=1,\dots, n_s$, coupled to the vector fields 
$F^\Lambda_{\mu\nu}$
in a non-minimal way through the real symmetric matrices 
$\mathcal{F}_{\Lambda\Sigma}(z,\bar{z})$, $\mathcal{R}_{\Lambda\Sigma}(z,\bar{z})$ 
and span a special K\"ahler manifold
with the metric $h_{i\bar{\jmath}}$. 
The scalar potential $V(z,\bar{z})$ originates from electric-magnetic  Fayet-Illiopulos terms;
 a consistent truncation with only one scalar field and a concrete potential was presented in \cite{Anabalon:2017yhv} (see, also, \cite{Anabalon:2013eaa, Lu:2014fpa, Faedo:2015jqa}). Despite the existence of a potential for the scalar field, one can consider a situation 
when the effective cosmological  constant vanishes at the boundary even if the self-interaction in the bulk does not \cite{Nucamendi:1995ex} -- exact asymptotically flat hairy black hole solutions 
with a non-trivial dilaton potential were obtained in \cite{Anabalon:2013qua}. 
 We observe that the moduli metric and coupling with the gauge field can be non-trivial. Depending of their form, the RN BH may, or may not, be obtained as a solution of the theory. For the dilatonic  coupling (\ref{dilaton}) 
and a trivial moduli metric, the limit $\phi \rightarrow 0$ does $not$ 
provide a consistent truncation and so RN BH is not a solution of the theory. However, this is not necessarily the case for any metric $h_{i\bar{\jmath}}$. 
Finding a concrete realisation of class II models in this context would be very promising. 

Amongst the several extension of the discussion herein, including in (\ref{action}) an axion-type term could be interesting, due to the high energy physics motivation for axions. Another obvious extension would be the consideration of solutions with less symmetry, either rotating solutions or solutions connected to zero modes with $\ell\neq 0$.

Finally, as a speculation, one can notice some analogy of the scalarised BHs and the AdS holographic duals of superconductors
(the $s-$wave case) \cite{Horowitz:2010gk}.
The general mechanism appears to be the following: for both asymptotics,
 the RN BH remains a solution of the full model. 
However, for some range of the parameters, the non-trivial coupling of the scalar field with the Maxwell field
 gives a tachyonic mass for the vacuum scalar perturbations around the RN BH, with the appearance of a scalar condensate. 
This implies the occurrence of a
branch of scalarised BHs, which are generically thermodynamically favoured over the RN configurations.
It would be interesting to further pursue this apparent parallelism in the Minkowskian case, and to investigate the possible
relevance of these aspects in providing `dual' descriptions to phenomena
observed in condensed matter physics.

\section*{Acknowledgements}
The work of D.A. has been funded by the Fondecyt Regular Grant 1161418. 
The work of  C.H., A.P., and E.R. is supported by the Funda\c{c}\~ao para a Ci\^encia e a Tecnologia (FCT) project UID/MAT/04106/2019 (CIDMA), by CENTRA (FCT) project UID/FIS/00099/2013, by national funds (OE), through FCT, I.P., in the scope of the framework contract foreseen in the numbers 4, 5 and 6
of the article 23, of the Decree-Law 57/2016, of August 29,
changed by Law 57/2017, of July 19. We acknowledge support  from the project PTDC/FIS-OUT/28407/2017.   
This work has also been supported by  the  European  Union's  Horizon  2020  research  and  innovation  (RISE) programmes H2020-MSCA-RISE-2015
Grant No.~StronGrHEP-690904 and H2020-MSCA-RISE-2017 Grant No.~FunFiCO-777740. 
A. P. is supported by the FCT grant PD/BD/142842/2018.  E. R.
gratefully acknowledges the support of the Alexander von Humboldt Foundation.
The authors would like to acknowledge networking support by the
COST Action CA16104.

 \appendix
\section{ Exact solutions  with a linear coupling}
\setcounter{equation}{0}
\renewcommand{\theequation}{A.\arabic{equation}}

\subsection{Purely electric  BHs}
Purely electric dilatonic solutions of~\eqref{action} with the dilatonic coupling~\eqref{dilaton} where first considered by Gibbons and Maeda~\cite{Gibbons:1987ps} and Garfinkle, Horowitz and Strominger~\cite{Garfinkle:1990qj}. The BH solution has the line element~(\ref{metric-generic}) with 
\begin{equation}  
\label{GHS-solution}  
a(r)^2=\frac{1}{c(r)^2}
=\left(1-\frac{r_+}{r} \right)\left(1-\frac{r_-}{r}\right)^{\frac{1-\alpha^2}{1+\alpha^2}}\ , \qquad
b(r)=r \left(1-\frac{r_-}{r}\right)^{\frac{ \alpha^2}{1+\alpha^2}}\ , 
\end{equation}
together with the Maxwell potential and dilaton field\footnote{Following the conventions in the work, we fix $\phi(\infty)=0$ for all solutions in the Appendix.}
\begin{equation}
A=\frac{Q}{r}dt \ , \qquad 
e^{2\phi}=\left(1-\frac{r_-}{r}\right)^{\frac{2\alpha}{1+\alpha^2}}\ .
\end{equation}
The two free parameters $r_{+}$, $r_{-}$ 
(with $r_-<r_+$)
 are related to the
ADM mass, $M$, and (total) electric charge, $Q$, by
\begin{eqnarray}   
  M = \frac{1}{2}
	\left[
	r_+ +\left(\frac{1-\alpha^2}{1+\alpha^2}\right)r_-
	\right] \ , \qquad 
	Q=\left(  
	\frac{r_-r_+}{1+\alpha^2}
	\right)^{\frac{1}{2}} \ .
\end{eqnarray}
For all $\alpha $, the
surface $r= r_H=r_+$ is the location of the (outer) event horizon,
with
\begin{eqnarray} 
A_H=4\pi r_+^2\left(1-\frac{r_-}{r_+} \right)^{\frac{2\alpha^2}{1+\alpha^2}},~~
T_H=\frac{1}{4\pi}\frac{1}{r_+-r_-}\left(1-\frac{r_-}{r_+}\right)^{\frac{2}{1+\alpha^2}} \ .
\end{eqnarray}
The extremal limit, which corresponds to the coincidence limit $r_- = r_+$, 
results in a singular solution (as can be seen $e.g.$ by evaluating the Kretschmann scalar).
In this limit, the area of the event horizon goes to zero for $\alpha \neq 0$.
The Hawking temperature, however, only goes to zero in the extremal limit for $\alpha<1$,
while for $\alpha=1$ it
approaches a constant, and for $\alpha>1$ it diverges.

The reduced quantities~\eqref{scale1} have the following exact expressions:
\begin{eqnarray}  
\label{GHS-solution-s1}  
\nonumber
q=\frac{2\sqrt{(1+\alpha^2)x}}{1+\alpha^2(1-x)+x},~~
 a_H=\frac{(1+\alpha^2)^2(1-x)^{\frac{2\alpha^2}{1+\alpha^2}}}
{(1+\alpha^2(1-x)+x)^2},~~
t_H=\frac{(1-x)^{\frac{1-\alpha^2}{1+\alpha^2}}(1+\alpha^2(1-x)+x)}
{1+\alpha^2}  \ ,
\end{eqnarray}
where
$0\leqslant x\leqslant 1$ is a parameter.

\subsection{Dyonic BHs}

\subsubsection{$\alpha=1$}
A dyonic dilatonic BH solution of~\eqref{action}, with the dilatonic coupling~\eqref{dilaton} and $\alpha=1$, was found in~\cite{Kallosh:1992ii}, and extensively discussed in the literature,
since it
 can be embedded in ${\cal N}=4$ supergravity.
Taking the form (\ref{metric-generic}), it has
\begin{eqnarray}
\label{b}
\phi= \frac{1}{2} \log\frac{(r+\Sigma)}{(r-\Sigma)}\, ,
\qquad
  a(r)^{2} = \frac{1}{c(r)^2}=\frac{(r-r_{+})(r-r_{-})}{(r^{2}-\Sigma^{2})}\, ,
\qquad
  b(r)^{2}=  r^{2}-\Sigma^{2} \ ,
	  \label{eq:alpha2_solution}
\end{eqnarray} 
where
\begin{equation} r_{\pm}=M\pm \sqrt{M^{2}+\Sigma^{2}- {Q}^{2}- {P}^{2}}\ ,
	\label{rhne}
\end{equation}
and the outer horizon is at $r_H=r_+$,
while
 $M,Q,P$ are the mass and electric and magnetic charges. $\Sigma$ corresponds to the
 scalar charge, which, however, is not an independent parameter (the hair is secondary):
\begin{equation}
  \Sigma=\frac{ {P}^{2}- {Q}^{2}}{2M}\ .
\end{equation}
The extremal limit of the above solution corresponds to  $r_+=r_-$,
in which case one finds 
 two relations between the charges
\begin{equation} 
  0 = M^{2}+\Sigma^{2}- {Q}^{2}- {P}^{2} \,\,\, 
	\Longrightarrow  \ \ (M+\Sigma)^2 -2 {P}^{2}=0 \qquad {\rm and} \qquad (M-\Sigma)^2 -2 {Q}^{2}=0 \ .
\end{equation}  

The horizon area and Hawking temperature of the solutions are 
\begin{eqnarray}  
A_H=4 \pi (2 M r_+ -P^2-Q^2)\ , \qquad T_H=\frac{1}{2\pi} \frac{r_+-M}{2M r_+-P^2-Q^2} \ .
\end{eqnarray}
 The expression of the reduced quantities is  more involved in this case:
\begin{eqnarray}  
\label{a=1-reduced}  
 a_H=\frac{1}{4} (2x-q^2)\ , \qquad
t_H=\frac{4(x-1)}{2x-q^2} \ ,
\end{eqnarray}
with  $x$ a parameter expressed in terms of $q$ as a solution of the equation (where $k=\frac{P}{Q}$)
\begin{eqnarray}  
\label{a=1-reduced2}  
q^4-\frac{4(1+k^2)^2}{(1-k^2)^2}(q^2+x(x-2))=0 \ .
\end{eqnarray}

\subsubsection{$\alpha=\sqrt{3}$}
A dyonic dilatonic BH solution of~\eqref{action}, with the dilatonic coupling~\eqref{dilaton} and $\alpha=\sqrt{3}$, was found in~\cite{Dobiasch:1981vh,Gibbons:1985ac}. This case arises from a suitable Kaluza-Klein reduction of a five-dimensional vacuum BH.
In the extremal limit,  
one obtains a non-BPS BH that can be embedded in ${\cal N}=2$
supergravity.

The generic solution can be written again in the form 
(\ref{metric-generic})
with
\begin{equation}
\label{kk1}
  a(r)^{2}=\frac{1}{c(r)^2} =\frac{(r-r_{+})(r-r_{-})}{\sqrt{AB}}\ , \qquad 
  b(r)^{2}= \sqrt{AB} \qquad {\rm and} \qquad  e^{4\phi(r)/\sqrt{3}} = \frac{A}{B} \ ,
\end{equation}
where
\begin{eqnarray}
  A= (r-r_{A_{+}})(r-r_{A_{-}})\ , \qquad
  B= (r-r_{B_{+}})(r-r_{B_{-}}) \ .
\end{eqnarray}
In the above relations one defines
\begin{equation}
  r_{\pm}=M\pm\sqrt{M^{2}+\Sigma^{2}-{P}^{2}-{Q}^{2}} \ ,
\end{equation}
where, again, the outer horizon is at $r_H=r_+$,
and
\begin{equation}
  r_{A_{\pm}}=\frac{1}{\sqrt{3}}\Sigma\pm {P}\sqrt{\frac{2\Sigma}
    {\Sigma-\sqrt{3}M}} \ , \qquad 
  r_{B_{\pm}}=-\frac{1}{\sqrt{3}}\Sigma\pm {Q}\sqrt{\frac{2\Sigma}
    {\Sigma+\sqrt{3}M}} \ .
\end{equation}

The solution possesses again three  parameters  $M,Q,P$
which fix the scalar charge $\Sigma$ via the equation
\begin{equation}
  \frac{2}{\sqrt{3}}\Sigma=\frac{ {Q}^{2}}{\sqrt{3}M+\Sigma}-
  \frac{ {P}^{2}}{\sqrt{3}M-\Sigma} \ ,
\end{equation}
while
the horizon area and the Hawking temperature are given by
\begin{eqnarray}
&&
A_H=4\pi\sqrt{(r_{+}-r_{A_{+}})(r_{+}-r_{A_{-}})
    (r_{+}-r_{B_{+}})(r_{+}-r_{B_{-}})}~,
		\\
\nonumber
&&
T_H=\frac{1}{4\pi}\frac{r_+-r_-}{\sqrt{(r_{+}-r_{A_{+}})(r_{+}-r_{A_{-}})
    (r_{+}-r_{B_{+}})(r_{+}-r_{B_{-}})}}~.	
\end{eqnarray} 

The corresponding expressions for $a_H$ and $t_H$
as a function of $q$ (and the ratio $P/Q$)
can be derived directly from the above relations;
however, they are too complicated to include  here.

 \begin{small}


 \end{small}

\end{document}